\documentclass[12pt,a4paper]{article}
\usepackage[utf8]{inputenc}
\usepackage[left=3cm,right=3cm,top=3cm,bottom=3cm]{geometry}
\usepackage{graphicx} 
\usepackage{xurl}
\usepackage{setspace}
\usepackage{ragged2e}
\usepackage{booktabs}
\usepackage{bbm}
\usepackage{amsmath}
\usepackage{amssymb}
\usepackage{placeins}
\usepackage{comment}
\usepackage[dvipsnames]{xcolor}
\usepackage{multirow}

\usepackage[dvipsnames]{xcolor}
\usepackage[colorlinks=true]{hyperref}
	\hypersetup{citecolor=MidnightBlue, anchorcolor=Black, urlcolor=MidnightBlue, linkcolor=MidnightBlue}
\usepackage[nottoc]{tocbibind}

\usepackage[style=apa]{biblatex}
\AtEveryBibitem{\clearfield{month}}
\AtEveryBibitem{\clearfield{day}}
\AtEveryBibitem{\clearlist{language}}
\AtEveryBibitem{\clearfield{note}}

\title{
Trapped in Transformative Agreements?\\ A Multifaceted Analysis of \textgreater1,000 Contracts}

\author{
Laura Rothfritz\thanks{Berlin School of Library and Information Science (IBI), Humboldt-Universität zu Berlin, Dorotheenstr.~26, 10117 Berlin, Germany \&
scidecode science consulting, Mühlenstraße 8a, 14167 Berlin, Germany.
Email: \href{mailto:laura.rothfritz@hu-berlin.de}{laura.rothfritz@hu-berlin.de}. \textbf{ORCiD}: \href{https://orcid.org/0000-0001-7525-0635}{0000-0001-7525-0635}}\\
{\small IBI, HU Berlin}
 \and 
 Ulrich Herb\thanks{Saarland University and State Library (SULB), Campus B1 1, 66123 Saarbrücken, Germany \& scidecode science consulting, Mühlenstraße 8a, 14167 Berlin, Germany. Email: \href{mailto:u.herb@sulb.uni-saarland.de}{u.herb@sulb.uni-saarland.de}. \textbf{ORCiD}: \href{https://orcid.org/0000-0002-3500-3119}{0000-0002-3500-3119}}\\
{\small SULB, Saarland U}
\and 
W.~Benedikt Schmal\thanks{Ilmenau University of Technology, Economic Theory Group, Ehrenbergstr.~29, 98693 Ilmenau, Germany \& KU Leuven University, Department for Management, Strategy, and Innovation (MSI), Naamsestraat 69, 3000 Leuven, Belgium. Email: \href{mailto:wolfgang-benedikt.schmal@tu-ilmenau.de}{wolfgang-benedikt.schmal@tu-ilmenau.de}. \textbf{ORCiD}: \href{https://orcid.org/0000-0003-2400-2468}{0000-0003-2400-2468}}\\
{\small TU Ilmenau}\\
{\small MSI, KU Leuven}
}
\date{\today}

\begin{document}
\begin{onehalfspace}

\maketitle  
\thispagestyle{empty}

\begin{abstract}
    \noindent Transformative agreements between academic publishers and research institutions are ubiquitous. The `Efficiency and Standards for Article Charges' (ESAC) Initiative lists more than 1,000 contracts in its database. We make use of this unique dataset by web-scraping the details of every contract to substantially expand the overview spreadsheet provided by the ESAC Initiative. Based on that hitherto unused data source, we combine qualitative and quantitative methods to conduct an in-depth analysis of the contract characteristics and the TA landscape. Our analysis demonstrates that research institutions seem to be `trapped' in transformative agreements. Instead of being a bridge towards a fully Open Access world, academia is stuck in the hybrid system. This endows the legacy (non-Open Access) publishing houses with substantial market power. It raises entry barriers, lowers competition, and increases costs for libraries and universities.
\end{abstract}
\noindent \textbf{Keywords: } \emph{Transformative Agreements, Academic Publishing, Open Access, Risk sharing, Elsevier, Springer Nature, Wiley}

\end{onehalfspace}
\newpage
\begin{doublespace}
    
\section{Introduction}
Transformative agreements (TA) were meant to be a bridge in academic publishing. Coming from a world built upon printing presses and mailed copies, the rapid spread of the Internet began to allow de facto cost-free dissemination and reproduction of academic knowledge. At the same time, large publishers bundled their journals in so-called `big deals' that sold, aside from flagship outlets, a large portfolio of journals, including a lot of intellectual by-catch. Exploiting the necessity of libraries to subscribe to the important journals from which researchers cannot shift away (\cite{bergstrom_free_2001}), research institutions were faced with rising costs for these big deals that were regularly based on paywalled journals (\cite{bergstrom_evaluating_2014}). 

While hard copies embodied a physical barrier to access, digital publishing allows effortless access for anyone with an internet connection. Free access to novel scientific knowledge allows for a faster spread of new information and enables researchers to have broader access to academic work. However, it does not necessarily fix systematic problems in science, e.g., the under-representation of lower-income countries (\cite{else_open_2024}). Long-standing, established publishers have been owning a large portfolio of subscription-based journals behind paywalls. Since fully changing this large network of academic outlets was not possible, TAs have been gradually put into place. Also known as `read and publish' or `publish and read' agreements, TAs are contracts between scholarly publishers and institutions designed to transition subscription-based journals to Open Access (OA) models. These agreements typically allow institutions to shift subscription costs to cover both reading and publishing fees (\cite{borrego_transformative_2021}).

TAs currently play a central role in the attempt to transform scholarly journal publications to OA. As the name suggests, the goal of TAs is the transformation of entire journals to full OA, the so-called `flipping' of a journal (\cite{solomon_longitudinal_2013}). As temporary agreements, TAs are thought to serve as an intermediary model until the flipping process is complete. Their adoption has grown significantly over the last few years, particularly in Europe and North America. The ESAC registry analyzed in this paper lists 1,075 contracts in August 2024, including agreements that have already been terminated.\footnote{See for a discussion on the limitations of this unqiue database subsection \ref{ssec.lim}.} Although there are general guidelines for TAs (ESAC, n.d.), covering copyright retention and the need for transparency, TAs vary in size between agreements by single institutions and large consortia. They also vary in other terms, e.g. their coverage of fully OA journals or just hybrid journals. 

In this paper, we examine virtually the entire body of TAs negotiated. Based on a mixed-method approach, we conduct an in-depth analysis of all past and present contracts. We ask to what extent academia may be locked or `trapped' in these contracts and whether TAs are the right way to change the academic publishing market for the better. The present paper provides a comprehensive examination of this contract landscape and its peculiarities. Furthermore, it makes use of the qualitative evaluation the registering institutions provide along with the contract details. The econometric analysis looks at changes over time and differences between contracts with the `Big 3' firms Elsevier, Springer Nature, Wiley, and other publishers. The qualitative analysis provides an overview of the most common themes mentioned in the comments submitted to the ESAC database by the institutions.

In doing so, this paper is the largest evaluation of TAs one decade after the first contracts became active. It addresses to which extent TAs fulfill the defined goals and their shortcomings. Our core result is that the academic side appears to be `trapped' in TAs in the sense that TAs do not seem to be transitory but the `new normal' in the publishing landscape.

The remainder of this paper is structured as follows. In Section \ref{sec.lit}, we review the existing literature on TAs and current issues of academic publishing. In Section \ref{sec.methdat}, we provide details on the data and methods we use. We present both our quantitative and qualitative results in Section \ref{sec.res}. We discuss the implications and limitations of our findings in Section \ref{sec.disc} before we briefly conclude in Section \ref{sec.conc}.

\section{Literature Review}
\label{sec.lit}
The evolution of OA publishing, particularly through TAs, reflects a significant shift towards more equitable access to research outputs. TAs are designed to facilitate the transition from subscription-based models to OA by allowing institutions to cover publication fees for their researchers while maintaining access to subscription content. This model has gained traction as a means to support the sustainability of both publishers and institutions, promoting wider dissemination of research findings (\cite{widding_beyond_2024}).

TAs were designed as a transitional mechanism to shift from traditional subscription models to OA, allowing institutions to repurpose their subscription expenditures to support OA publishing (\cite{borrego_transformative_2021}). The development of TAs gained momentum in the late 2010s, with key stakeholders such as research institutions, consortia, and publishers recognizing the necessity of restructuring their financial models to accommodate OA (\cite{inchcoombe_transforming_2022}). Early agreements, such as the `Read and Publish' models, were forged in response to growing dissatisfaction with subscription costs and the desire for increased access to research publications (\cite{anders_read_2021}). These agreements were also motivated by the need to comply with OA mandates from funding bodies, which required that publicly funded research be made freely available. The context in which these agreements evolved was shaped by the broader movement towards OA, characterized by the principles outlined in initiatives like Plan S in Europe, which further accelerated the adoption of TAs (\cite{quaderi_plan_2019}).

TAs can be broadly categorized into two primary types: \textit{“Read and Publish” (RAP)} and \textit{“Publish and Read” (PAR)} agreements (\cite{borrego_transformative_2021}). A RAP agreement combines subscription access with OA publishing. Under this model, institutions pay a fee that covers both access to the publisher’s subscription journals and the costs associated with publishing articles by affiliated authors in OA. This model is beneficial for institutions seeking to manage their expenditures while increasing the OA output of their researchers. The costs of RAP agreements are essentially based on the access costs (i.e. the former subscription costs), whereas the costs of PAR agreements are based on the publication (and therefore OA) costs, with access to the publisher's subscription content included as a secondary benefit. Both models aim to repurpose the funds traditionally spent on subscriptions to support OA, but only PAR contracts are considered to be compliant with the OA transformation (\cite{machovec_strategies_2019}).

In addition to these, other hybrid modes of TAs have emerged, tailored to specific institutional needs and publishing landscapes. These include `Subscribe to Open' (S2O) models, where a journal remains subscription-based but converts to OA if sufficient subscription revenue is secured from institutions. If the revenue target is met, the journal becomes OA for that year; otherwise, it reverts to a subscription model (\cite{bosshart_open_2022}). Another, retiring mode, are `Offsetting Agreements', which aims to offset the costs of OA publishing by reducing subscription fees as OA output increases. These agreements are often customized and negotiated on a case-by-case basis, considering factors such as the institution's publishing output, the specific needs of the scholarly community, and the financial viability of the publisher (\cite{earney_offsetting_2017}).

TAs can enhance collaboration between libraries and publishers, fostering a more integrated approach to scholarly communication (\cite{bakker_impact_2024}). However, challenges remain, including the need for transparency in pricing and the potential for inequities between institutions with varying financial capabilities (\cite{finnie_another_2023}). While TAs are a promising step, they are not a panacea; ongoing evaluation is necessary to ensure that they effectively meet the objectives of expanding access and supporting diverse research outputs (\cite{hoogendoorn_scaling_2023}).

Research on the impact of TAs has shown a complex landscape of benefits and challenges. Bakker et al. highlight that TAs can enhance access to scholarly content, particularly for institutions with limited budgets, thereby promoting equity in research dissemination. In addition, TAs promote collaboration between institutions and publishers, fostering a shared commitment to OA principles and improving the visibility of research outputs (\cite{bakker_impact_2024}). By consolidating costs for both reading and publishing into a single agreement, TAs simplify budget management for libraries. This can make financial planning more predictable and help libraries maximize the value of their investments in scholarly communications (\cite{campbell_how_2022}). This predictability does not really apply when payments are calculated purely on publication output, and it is considered that the volume of articles produced annually grows on average by 3.8\% (\cite{white_publications_2019}). Universities often have mechanisms in place to allocate the costs of OA activities to disciplines and their departments or faculties. However, these are also confronted with volatility, e.g. due to the surge in the output of virology publications in the wake of the coronavirus pandemic (\cite{sellers_impact_2023}).

Negotiating and implementing TAs can be complex and resource-intensive for libraries. The process requires significant administrative effort, including checking the eligibility of submissions, tracking OA publishing outputs, managing APCs and invoices, and ensuring compliance with the terms of the agreements (\cite{parmhed_transformative_2023}). The shift towards TAs has also transformed the roles of libraries and librarians, requiring them to take on new responsibilities related to OA publishing support, financial management, and contract negotiation. This can lead to challenges in adapting to these new roles and ensuring that library staff have the necessary skills and resources (\cite{campbell_how_2022}). Jahn points out that the transition to TAs requires careful management to avoid possible disruptions in publishing workflows and to ensure that all stakeholders are adequately supported during the transition (\cite{jahn_how_2024}).

Generally, TAs may increase OA publishing, but function mostly by shifting payment streams from subscriptions to RAP or PAR fees. This allows publishers to secure steady income streams by converting subscription fees into funds that cover both content access and OA publishing costs (\cite{butler_oligopolys_2023}). TAs may also reinforce the dominant position of dominant publishers (e.g., the `Big 3' Elsevier, Springer Nature and Wiley). While \cite{haucap_impact_2021} identify a stronger market position of Springer Nature and Wiley in the field of chemistry due to the German DEAL TAs, \cite{schmal_how_2024} finds in a multidisciplinary follow-up study, the more nuanced finding of a `Matthew' effect (\cite{merton_matthew_1968}). In the context of publishers, this means that TAs tend to strengthen publishers, especially in those disciplines in which they already held a strong position beforehand. This position is reinforced by the fact that, unlike other publishers, legacy publishers benefit from a reputation bonus through the more or less guaranteed indexation of their journals in exclusive (impact) databases, e.g., Web of Science or Scopus (\cite{van_bellen_oligopoly_2024}), which provides a further incentive to publish findings in their outlets. 

By locking in large institutional agreements, these publishers can maintain or even increase their market dominance, as institutions feel pressured to include access to a broad range of journals in these agreements due to their essential academic content. TAs may lead to inflated costs for academic institutions. The `must stock' nature of journals published by the `Big 3' forces institutions to continue paying high fees under the guise of transitioning to OA.\footnote{While it is plausible that libraries need to provide their researchers with access to (up to now) large amounts of paywalled literature published by the large publishers, examples such cancellation of contracts with Elsevier by German and Swedish universities (\cite{else_dutch_2018}) have shown that accessability as the 'read' component of TAs may not be the most important part (\cite{fraser_no_2023}).  However, it could be that this was an exemption rather than the norm in the sense that large publishers would take additional steps against so-called shadow libraries (\cite{maddi_culture_2023}) in case reading access would be canceled on a large scale.} This creates a scenario where institutions might not see a significant cost reduction despite the shift from subscription to OA. (\cite{schmal_must_2024}). Schmal also emphasizes that while TAs facilitate OA publishing, they can also lead to increased costs for institutions, raising concerns about sustainability and long-term financial implications (\cite{schmal_how_2024}). In addition, studies have found that TAs might reinforce disciplinary, institutional, and geographic priorities. Fields that traditionally have fewer resources or lower publication volumes may not benefit equally from TAs, leading to an uneven distribution of OA publications. This could exacerbate existing inequities between well-funded and less-funded disciplines (\cite{parmhed_transformative_2023, bakker_impact_2024}).  Institutions in wealthier regions with the resources to negotiate favorable TAs are more likely to see an increase in OA publications, while those in less affluent regions may be left behind, thus widening the global knowledge gap (\cite{anderson_how_2022}; \cite{moskovkin_transformative_2022}; \cite{momeni_which_2023}).

TAs were initially conceived as temporary mechanisms aimed at facilitating the transition of subscription-based journals to full OA. The underlying expectation was that, over time, these agreements would lead to the `flipping' of journals, permanently establishing OA as the norm (\cite{jisc_transformative_nodate}; \cite{coalition_s_criteria_nodate}). However, as recent data suggests, this objective is far from being realized. According to a 2023 analysis by Coalition S (\cite{kiley_transformative_2024}), many journals under TAs are not flipping to OA at the anticipated rate. Similarly, other studies have suggested that the effects of TA are scaled toward transitioning closed access journals to hybrid OA for researchers and not towards transitioning journals to fully OA for publishers (\cite{jahn_how_2024}; \cite{brayman_review_2024}). This stagnation creates a scenario in which institutions and consortia find themselves perpetually engaged in TA negotiations rather than moving towards a sustainable OA future (\cite{green_is_2019}; \cite{fenter_it_2022}). Instead of serving as a bridge to full OA, TAs risk becoming a long-term commitment, trapping institutions in a cycle of ongoing and potentially costly negotiations with no clear end in sight.

For this study, we used the \textit{Transformative Agreements Registry} provided by the ESAC initiative\footnote{See: \url{https://esac-initiative.org/}. 
} Previous studies on TAs also used this database. For example, \cite{borrego_transformative_2021} used the database to extract 36 TAs and analyzed their full-text version in depth. \cite{szprot_transformative_2021} provided a legal analysis of TAs found in the ESAC database. \cite{marcaccio_transforming_2022} analyzed a sample of 14 agreements based on the ESAC database. A first formal statistical analysis of the database itself was done by \cite{moskovkin_transformative_2022}. The authors used the publicly available version (Excel spreadsheet) of the data and analyzed it to show a significant increase in TA adoption between 2020 and 2021, predominantly in wealthier countries, such as the Netherlands, the UK, Germany, and the United States (\cite{moskovkin_transformative_2022}). Recently, \cite{bakker_impact_2024} studied the impact of TAs on publication patterns based on a sample of 74 TAs from the ESAC database enriched with publication data for journals included in the TA. The publicly available TA Data dump\footnote{See \url{https://journalcheckertool.org/transformative-agreements/}, last checked 9 August 2024.} identifies TAs according to their ESAC ID. Jahn used this dump and the ESAC registry in conjunction with the publication metadata database to study the development of hybrid OA (\cite{jahn_how_2024}). Eventually, \cite{mccabe_open_2024} utilize information from this database for a more general discussion of open access policies. Most studies focused on active TAs. For our study, it did not matter whether TAs were still active or not since we were interested in developments over time. Therefore, we used the full sample available to us via ESAC at the time of writing this paper: 1,075 TAs (9 August 2024).

\section{Methods and Data}
\label{sec.methdat}
\noindent \textbf{Data:} We base our analysis on the ESAC database, which aggregates information on TAs in a standardized way. It is part of the ESAC initiative, which is organized by the Max Planck Digital Library and run by academic library practitioners in Europe.\footnote{See also \url{https://esac-initiative.org/about/about-esac/}, last checked on 27 August 2024.} To the best of our knowledge, the ESAC registry for TAs is globally the largest database on TA contracts that is publicly available. Although the initiative provides an easy-to-use overview as a download option, we gathered all the details that are collected for each contract. In particular, we scraped the list of contracts and added additional records that are listed on a separate `contract details' page to every contract entry (\cite{rothfritz_trapped_2024}). Although an overview is available directly on the ESAC website, to the best of our knowledge, we are the first to amend this list with contract-specific details. Figure \ref{fig:esac_screenshots} in the appendix illustrates the differences. Of course, the quality of our analysis depends on the quality of the registry entries in the ESAC database.\footnote{See also our discussion on limitations in subsection \ref{ssec.lim}.} As long as errors are idiosyncratic and not systematic, any bias should be negligible. All data and code are openly available for reuse (see Data Availability Statement).\\

\noindent \textbf{Quantitative Methodology:} In the quantitative part of the analysis, we made use of the large quantity of numerical information in the ESAC database. In addition to descriptive statistics and comparisons based on averages separated by selected criteria, we used regression analysis to elicit correlational evidence on the change in the number of concluded TAs over time. Furthermore, we set up repeated cross-section fixed effect regressions to investigate the dynamic nature of TAs. Here, we empirically analyze whether the size of an initial agreement increases the likelihood that this contract will receive a renewal. This will shed light on the presence of an additional Matthew effect in the sense that larger contracts are more likely to be prolonged, which is likely to benefit particularly the larger publishing houses.

We included fixed effects for the publisher, the starting year of a TA, and the country where the negotiating library consortium is based. We did this to net out noise or biases created by differences in the three named dimensions. For example, countries with more higher education institutions might negotiate larger TAs. To ensure that such underlying variation did not accidentally drive our results when looking at something else, we include these fixed effects, such that we can isolate effects \emph{within} some dimension, for example, netting out variation \emph{across} countries.
To make it more convenient for the reader to interpret the regression results, we provide the exact regression equations directly ahead of the results tables in Section \ref{ssec.emp}, where we present our quantitative findings. \\

\begin{table}[ht!]
    \centering
    \footnotesize
    \begin{tabular}{p{6.5cm} p{3cm} p{5cm}}
    \toprule
       Theme  & Method & Codes   \\ 
       \midrule
        Capping of agreement & Deductive coding &  YES\\
         & & NO \\
         & & N/A \\ \midrule
        Comments on risk sharing & Inductive coding & Gold OA not included \\
         & & Roll-over possible\\
         & & No roll-over possible \\
         & & APC discounts \\
         & & No APC discounts \\
         & & Refunds possible \\
         & & No refunds \\ 
         & & Other \\ \midrule
        Comments on workflows positive & Inductive coding &  CCC Rightslink\\
        & & Dashboards \\
        & & Automated Workflow\\
        & & Other \\ \midrule
        Comments on workflows negative & Inductive coding &  Complicated\\
        & & Manual workflow\\
        & & No monitoring of workflow possible \\
        & & Other \\ \midrule
        Overall Assessment & Deductive coding & Generally positive\\
        & & Generally negative \\
        & & Too early to assess\\ 
         & & N/A\\ \midrule
        Comments on overall assessment negative & Inductive coding &  Author Engagement \\
        & & Contract Terms \\
        & & Financial Issues \\
        & & OA Exploitation \\
        & & Publishing Services \\
        & & Other \\ \midrule
        Comments on overall assessment positive & Inductive coding & Transition and Transformation \\
        & & Sustainable regarding costs \\
        & & No cap \\
        & & Good workflows \\
        & & Transparency \\
        & & Other \\
        \bottomrule
    \end{tabular}
    \caption{Deductive \& inductive codes for the qualitative analysis of free-text answers.}
    \label{tab:codes}
\end{table}

\noindent \textbf{Qualitative Methodology} ESAC collects its data via an online form, which contains both single-choice and free-text answer opportunities. Our dataset contained answers to all 26 questions. None of the free-text answers are mandatory to complete the form, meaning that our dataset does not include free-text answers for all agreements. The data was cleaned using OpenRefine Version 3.7.2 and thematically coded using a spreadsheet. Inductive codes were converged in OpenRefine through faceting and clustering. The textual data were sequentially coded using a high-level thematic analysis (\cite{braun_using_2006}; \cite{kiger_thematic_2020}) based on the developed themes, shown in Table \ref{tab:codes}.
 
Deductive codes were used for relatively dichotomous cases, such as \textit{capping/no capping}, and \textit{positive/negative} valences. Deductively coded free text was also coded inductively and in more depth in later rounds of coding. 
Inductive codes were developed based on the free text fields provided by the online form. All comments were read and common themes were identified and coded. Two authors coded it to ensure inter-coder reliability. It was possible to code sections of text with multiple codes. The analysis focused on quantities of theme occurrences and inductive codes were aggregated to a common form over several rounds of coding.
\FloatBarrier

\section{Results}
\label{sec.res}
\vspace{-4mm}
\subsection{Quantitative Analysis}
\label{ssec.emp}
\begin{figure}[htbp]
    \centering
    \includegraphics[width=.6\linewidth]{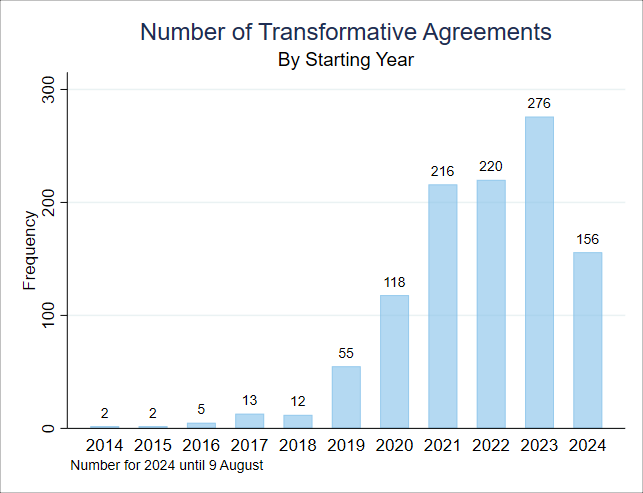}
    \caption{Aggregate Number of TAs by Year}
    \label{fig:ta_year}
\end{figure}

\noindent \textbf{Agreements per Publisher:} The 1,075 TAs in the ESAC database have been negotiated with 65 publishers in various disciplines. Looking at the evolution over time, presented in Figure \ref{fig:ta_year}, it becomes obvious that TAs continue to gain traction. After a very slow beginning in the years 2014 - 2018, we observed rapid growth since 2019. In 2023, the number of newly starting TAs peaked -- so far -- with 276 TAs beginning. This includes prolongations and renewals that are listed as separate contracts in the ESAC database. Until August 2024, we record 156 contracts. It is not clear whether this means that TAs lose momentum or whether a substantial amount of contracts has not been registered yet.\footnote{Given that a listing in the ESAC database is voluntary, this may not take place immediately after a contract was concluded or became active.}

\begin{figure}[htbp]
    \centering
    \includegraphics[width=.49\linewidth]{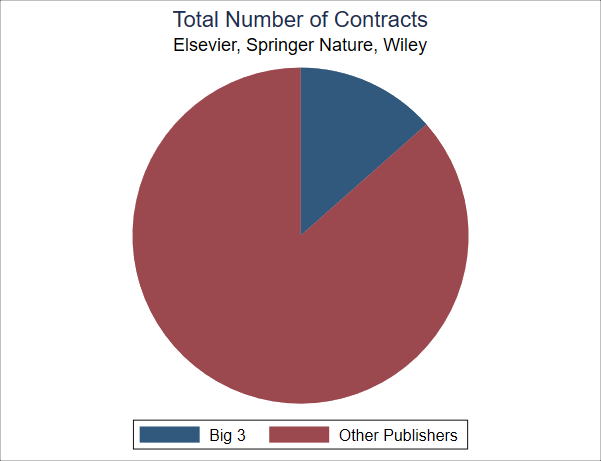}
    \includegraphics    [width=.49\linewidth]{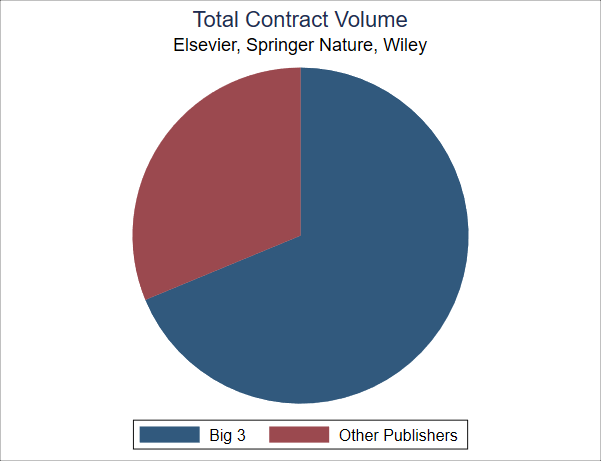}
    \caption{Aggregate Number and Volume of TAs of the `Big 3' publishers}
    \label{fig:big3_vol1}
\end{figure}

However, the number of contracts per publisher varies a lot, as Table \ref{tab.publ_all} in the appendix demonstrates. While more than half of the publishers (36) negotiated less than ten contracts, Cambridge University Press alone successfully negotiated 75 TAs. The `big 3' publishers concluded 53 (Springer Nature), 51 (Wiley), and 41 (Elsevier) agreements, respectively. Although they make up 13.5\% of all contracts, their volume is enormously larger. The number of planned annual publications accounts for approximately 3/4 (73.3\%). Only one out of eight agreements have been negotiated with the `Big 3', but three out of four publications covered by TAs globally appear in journals of the leading commercial publishing houses Elsevier, Springer Nature, and Wiley. Figure \ref{fig:big3_vol1} illustrates this.\\

\noindent \textbf{Geographical Dispersion and Size:} 
Of the 1,075 agreements in the registry, seven were organized via the network Electronic Information for Libraries (EIFL), the others were negotiated by consortia or individual universities. The variation across agreements is quite large. While many TAs concluded between a publisher and a single university may include only very few publications, others like the German DEAL contracts or the British contracts organized by JISC have five-figure volumes per year. If the EIFL contracts are categorized by the individual countries involved, it can be seen that, according to ESAC, 76 countries have concluded agreements, 27 of which have no national agreements and are solely contracting partners through the multinational EIFL consortia. 

Looking only at agreements with a start date between 2018 and 2020, the following pattern emerges. There are 184 national consortia and only one multinational EIFL consortium. Of the agreements starting between 2021 and 2023, three were multinational EIFL consortia and 709 national. Here, 32 nations participated exclusively in agreements through EIFL contracts. Consequently, 185 consortia were launched between 2018 and 2020 and 712 between 2021 and 2023. Contracts with the start year of 2018-2020 allowed organizations from 210 countries to participate in the agreements and those with the start year of 2021-2023 allowed organizations from 795 countries. This indicates that a growing number of countries only benefit from TAs through international cooperation, which may be due to a lack of negotiating power, too little commercial incentive on the publisher's part for national deals, or too few resources for complex negotiations with publishers. It is also noticeable that not only the number of TAs is increasing but also the number of countries that want to use them as an option to increase the OA output.

\begin{figure}[htbp]
    \centering
    \includegraphics[width=0.9\linewidth]{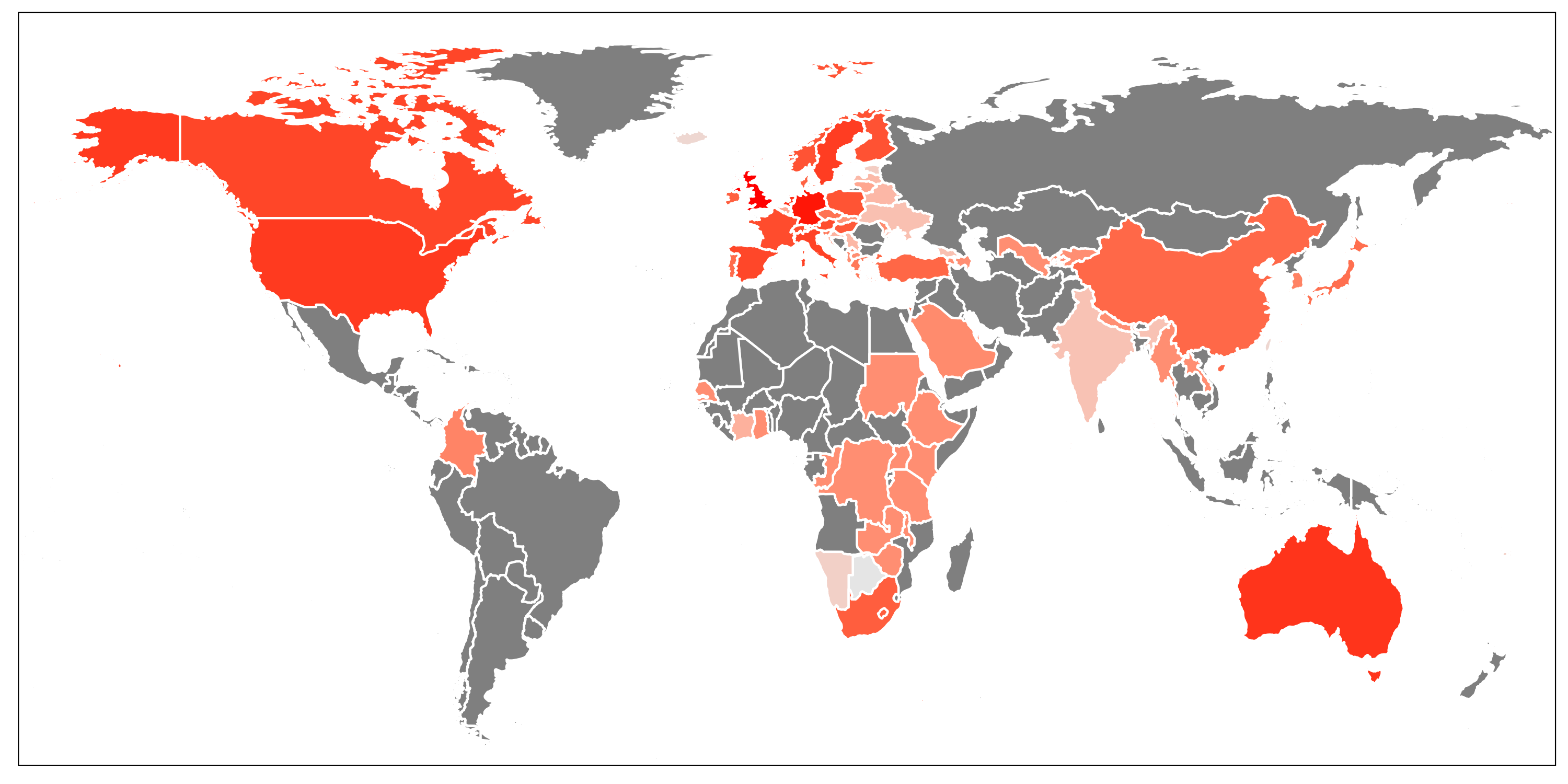} 
    \caption{Map of accumulated TA volumes by country. A higher color intensity implies a larger sum of annual TA publications.}
    \label{fig:country_vol1}
\end{figure}

Two measures are especially relevant: the total number of contracts per country, which addresses how widespread TAs are in the academic landscape, and the aggregated volume of these contracts. The latter captures the economic relevance of these contracts, as the remuneration for PAR agreements is usually based on pay-per-article fees. Figure \ref{fig:country_vol1} shows a world map that plots differences across countries in terms of annual publications. As already shown, TAs are a Eurocentric topic. Six countries concluded more than every second TA globally. Besides the United States, these are Germany, the United Kingdom, the Netherlands, Austria, Hungary, and Sweden, which jointly account for 45.95\% of all TAs concluded in the past and present (see also Table \ref{tab.coun_all} in the appendix). 

In Figure \ref{fig:country_vol_small} in the appendix, we show pie charts that show that the volume of agreements correlates with the number of agreements. This is also robust when excluding contracts with low publication volumes.\\

\begin{figure}[htbp]
    \centering
    \includegraphics[width=.49\linewidth]{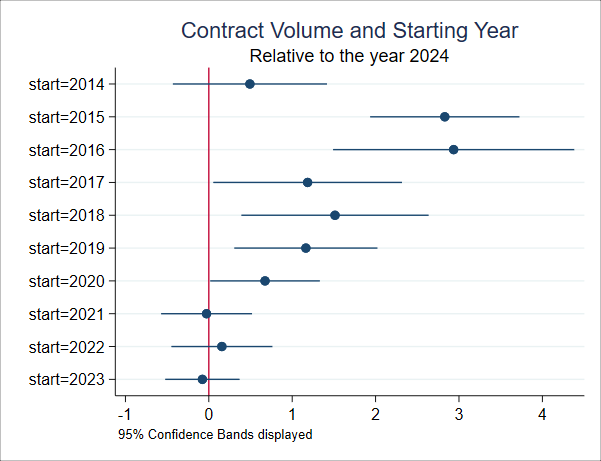}\\
    \caption{Relationship between Size and Starting Year of TAs. Coefficients based on the regression displays in eq.~\ref{eq1}. Heteroskedasticity robust standard errors clustered on the publisher level. 95\% confidence bands displayed.}
    \label{fig:yearsize_reg}
\end{figure}

\noindent \textbf{Evolution over time:} 
To detect changes over time, we regress the logarithmic size of the TAs on the vector $\theta_t$ of categorical time indicator variables for the years 2014 - 2023. We omit the coefficient for 2024 as the reference category. This is displayed in equation \ref{eq1}:
\begin{align}
    log(size)_{it} = \beta_0 + \beta_1\theta_t + \epsilon_{it}\quad \forall\:i \in [1, 1075] \:and\: t \in [2014, 2024)
    \label{eq1}
\end{align}
We cluster the standard errors on the publisher level because it is likely that the size of the TAs is correlated with the publishers as, in general, publishers like Springer Nature, Wiley, or Elsevier are more likely to negotiate larger contract volumes than small university presses or niche publishers.  Figure \ref{fig:yearsize_reg} presents the coefficients. As one can see, the contract size is indifferent for the years 2021-2023 relative to 2024 (up to August '24). However, significant positive estimates are observed for the years 2015 - 2020. Put differently, the average contract size of TAs that became active in the respective year has been significantly higher relative to those TAs that began in 2024. 

\FloatBarrier

While contract sizes have been higher in earlier years compared to 2024, we generally find a positive statistical relationship between the annual publication volume and its length. These results are widely robust to including fixed effects. Table \ref{tab:size_length} below emphasizes this. The results follow the following regression equation: 
\begin{align}
    log(size)_{iptc} = \beta_0 + \beta_1 \lambda + \beta_2\theta_p + \beta_3\theta_t + \beta_4\theta_c + \epsilon_{iptc}\quad 
    \label{eq2}
\end{align}
Again, $\beta_0$ represents the constant. The variable $\lambda$ captures the length of the TA or its duration, the independent variable of interest. Step-wise, we include fixed effects that shall net out effects caused by publishers, time, or country of the consortium. By that, we ensure that our results are not driven by differences across one of the three dimensions. To ensure that this relationship is not driven by many small contracts, where slight variation in size is rather negligible from a `big picture' perspective, we exclude in two steps TAs $<10$ and TAs $<100$ annual publications. The results are shown in Table \ref{tab:big3_length2} in the appendix and are similar to the results for the full sample presented in Table \ref{tab:size_length}. 

\begin{table}[ht!]
\small
    \centering
    \begin{tabular}{lrrrr}
    \toprule
    \multicolumn{5}{c}{\textbf{\emph{Dependent variable: TA volume}} (log)} \\
    \midrule
      &     \multicolumn{1}{c}{(1)}   &       \multicolumn{1}{c}{(2)}   &       \multicolumn{1}{c}{(3)}  &     \multicolumn{1}{c}{(4)}   \\
     Coefficient            & \multicolumn{1}{c}{OLS} & \multicolumn{1}{c}{OLS} & \multicolumn{1}{c}{OLS} & \multicolumn{1}{c}{OLS} \\
 \midrule
log(TA-Duration)    &       1.286***&       0.715***&       0.701***&       0.514***\\
                    &      (0.24)   &      (0.11)   &      (0.11)   &      (0.15)   \\
Constant            &      -4.590***&      -1.009   &      -1.338   &      -2.257   \\
                    &      (1.49)   &      (0.80)   &      (1.17)   &      (1.56)   \\
\midrule
\multicolumn{5}{c}{\textbf{\emph{Fixed Effects}}} \\
\midrule
Publisher & - & \checkmark & \checkmark & \checkmark \\
Start year & - & - & \checkmark & \checkmark \\
 Country & - & - & - & \checkmark\\
\midrule
R$^2$               &       0.088   &       0.612   &       0.614   &       0.734   \\
BIC                 &    4630.370   &    3704.328   &    3762.444   &    3598.523   \\
N                   &        1074   &        1074   &        1074   &        1074   \\
\bottomrule
    \end{tabular}\\
    {\footnotesize $* p<0.10, ^{**} p<0.05, ^{***} p<0.01$}
    \caption{Regression results for the relationship between TA volume and its length. Estimation method: OLS following eq.~(\ref{eq2}). Dependent variable: log(TA Volume). 95\% standard errors in brackets below, heteroskedasticity-robust and clustered on the publisher level. Results for the samples in which we exclude TAs $<10$ and TAs $<100$ annual publications are shown in Table \ref{tab:big3_length2} in the appendix.}
    \label{tab:size_length}
\end{table}

The significantly positive coefficient for the duration of transformative agreements indicates that there exists a positive relationship between contract size and contract length, even if we take into account differences across publishing houses, time, and the country in which the consortium is based (as statistically done by including the respective fixed effects). 

\begin{table}[htbp]
\footnotesize
    \centering
    \begin{tabular}{rccccc}
\toprule
\multicolumn{6}{c}{\textbf{\emph{Average length of TAs}}} \\
 & Number & Mean & Std.~dev. & Min & Max \\
\midrule
Overall &   1,075 &  2.47 &   1.05 & 0.33 &   8 \\
 $\mathbbm{1}_{big\:3}$ = 0 &   930  &  2.38  &   1.02  & 0.33 &   8 \\
 $\mathbbm{1}_{big\:3}$ = 1 &   145  &  3.01  &  1.04  & 0.50 &   6 \\
\bottomrule 
    \end{tabular}
    \caption{Summary statistics for the length (in years) of TAs}
    \label{tab:length_avg}
\end{table}

Digging deeper into this relationship, we also examine differences between TAs concluded with Elsevier, Springer Nature, and Wiley -- the big 3 publishing houses -- and TAs with any other publisher. Table \ref{tab:length_avg} displays summary statistics on the average length of all 1,075 TAs as well as distinguished by whether the `Big 3' negotiated them or any other publishing house. While the average length of TAs concluded with `other' publishers amounts to 2.38 years, contracts with Elsevier, Springer, or Wiley last for 3.01 years on average. This is backed by the regression results shown in Table \ref{tab:big3_length} in the appendix, where we also net out variation in country and/or time. It reports a significantly positive estimate for a contract concluded with a `Big 3' publisher on the length of this contract. 

\begin{table}[ht!]
\footnotesize
    \begin{center}
    \begin{tabular}{rcccccc}
\toprule
\multicolumn{7}{c}{\textbf{\emph{Average length of TAs}}} \\
\midrule
Country & $\mathbbm{1}_{big\:3}$ & Number & Mean & Std.~dev. & Min & Max \\
\midrule
 \multirow{2}{*}{Germany} & $\mathbbm{1}_{big\:3}$ = 0 &  118  &  2.482 &   1.120  &  1 &  8 \\
&  $\mathbbm{1}_{big\:3}$ = 1 &   8  &  4.191  & 1.148  &   2 & 5.34 \\
 \midrule
 \multirow{2}{*}{The Netherlands} & $\mathbbm{1}_{big\:3}$ = 0 & 109 & 2.387 & 1 & 0.915 & 5.17 \\
& $\mathbbm{1}_{big\:3}$ = 1 & 9 &  3.447 &  1.686 & 0.504 & 6 \\
 \midrule
 \multirow{2}{*}{United Kingdom} & $\mathbbm{1}_{big\:3}$ = 0 & 91 & 2.055 &  0.752 & 1 & 4 \\
& $\mathbbm{1}_{big\:3}$ = 1 & 5  &  3.203  &  0.837 &  2 & 4 \\
 \midrule
 \multirow{2}{*}{Austria} & $\mathbbm{1}_{big\:3}$ = 0 &  54  &  2.540 & 1 &  1 & 6 \\
& $\mathbbm{1}_{big\:3}$ = 1 &  9  &  2.891 & 0.334  &  2 & 3 \\
 \midrule
 \multirow{2}{*}{Hungary} & $\mathbbm{1}_{big\:3}$ = 0 & 40  &  1.176   & 0.550 & 1  & 3 \\
& $\mathbbm{1}_{big\:3}$ = 1 &  9  & 1.612  &  0.930  &  1 &  3  \\
 \midrule 
   \multirow{2}{*}{Other European Countries} & $\mathbbm{1}_{big\:3}$ = 0 & 301 & 2.471 & 1.018 & 0.334 & 5 \\
& $\mathbbm{1}_{big\:3}$ = 1 &  64 & 3.132 & 0.846 & 0.732 & 5 \\  
  \midrule
 \multirow{2}{*}{United States} & $\mathbbm{1}_{big\:3}$ = 0 &   72  &   2.653  &  0.871  & 0.723 &  5 \\
& $\mathbbm{1}_{big\:3}$ = 1 & 11  &  2.517  &  1.616 &  0.915 &  6 \\
 \midrule
 \multirow{2}{*}{China} & $\mathbbm{1}_{big\:3}$ = 0 &  5 &  2.602  & 1.674 & 1  &  5\\
& $\mathbbm{1}_{big\:3}$ = 1 &  0 & - & - & - & - \\
 \midrule
 \multirow{2}{*}{Other Countries} & $\mathbbm{1}_{big\:3}$ = 0 &   140 & 2.460 & 0.951  &  1  & 6 \\
& $\mathbbm{1}_{big\:3}$ = 1 &  30 & 2.907 & 0.465 & 1.888  &  4 \\
\bottomrule 
    \end{tabular}
        \end{center}
    \caption{Summary statistics for the length (in years) of TAs. Note that `European' here has a geographical and not a political meaning, i.e., it captures the entire continent, not only the European Union. Turkey is not part of that group.}
    \label{tab:length_avg2}
\end{table}
\FloatBarrier

The difference can have a substantial impact on business, as longer contracts imply longer payment streams generated by TAs, especially since academic publishers generally only have two to three years of financial planning stability (\cite{aspesi_landscape_2019}). Even though the exact number of submissions and publications, respectively, are uncertain, the `Big 3' publishers have longer time spans for planning and organizing their businesses - especially in the case of RAP agreements. Of course, the number of agreements concluded with publishers aside from the `Big 3' exceeds by far those concluded with the leading publishers (as also shown in Table \ref{tab:share_Big3_Country3} in the Appendix. However, the difference in time is notable. Looking closer at heterogeneity across countries -- as done in Table \ref{tab:length_avg2} -- it is interesting to see that especially the Netherlands, the United Kingdom, and Germany have stark differences in the average contract length, approximately 1, 1.2, and 1.5 years.\\
\FloatBarrier

\noindent \textbf{Contract Renewals:} Important for our headline question to which extent research institutions may be `trapped' in transformative agreements is the question of what determines the renewal of contracts. In theory, TAs should transform the large group of hybrid journals that offer subscriptions and, additionally, open access for purchase into gold open access journals that no longer sell subscriptions. Contract renewals do not say anything about how many journals `flipped.' In contradiction, renewals may be necessary as long as many journals stick to the traditional hybrid business model. Hence, renewals should not be attractive to research institutions if the ultimate goal was a full transformation to gold open access. 

We look at the volume of TAs as driver for renewals. A significantly positive statistical relationship would imply that larger contracts have a higher likelihood to be renewed. It would mean two things: Larger publishers (which arguably negotiate larger contracts in the first place) have higher chances to get renewals, which may be beneficial for their business. Furthermore, it implies a `Matthew' effect that larger contracts may get more easily extensions. To investigate this question, we use the following regression design:
\begin{equation}
        \mathbbm{1}_{Renew} = \beta_0 + \beta_1 \lambda_i + \beta_2\sigma_i + \beta_3\theta_p  +\beta_3\theta_t + \beta_4\theta_c + \epsilon_{iptc}\quad 
    \label{eq3}
\end{equation}
We define a dependent binary dummy $ \mathbbm{1}_{Renew} $, that turns one if a follow-up contract exists and remains zero otherwise. We regress it on contract length ($\lambda_i$) and size ($\sigma_i$) as well as on fixed effects for the starting year, publisher, and country of the negotiating consortium/institution. Due to the binary setting, we mainly rely on probit regressions and only use ordinary least squares (OLS) as a robustness check. 

\begin{table}[ht!]
\footnotesize
    \begin{center}
    \begin{tabular}{lrrrr|r}
    \toprule
    \multicolumn{6}{c}{\textbf{\emph{Dependent variable: New TAs getting a renewal}}} \\
    \midrule
                    &       \multicolumn{1}{c}{(1)}   &       \multicolumn{1}{c}{(2)}   &       \multicolumn{1}{c}{(3)}  &       \multicolumn{1}{c}{(4)} & \multicolumn{1}{c}{(5)} \\
   Coefficient            & \multicolumn{1}{c}{Probit} & \multicolumn{1}{c}{Probit} & \multicolumn{1}{c}{Probit} & \multicolumn{1}{c}{Probit} & \multicolumn{1}{c}{OLS} \\
                    \midrule
log(TA Size)      &       0.058** &       0.109***&       0.080*  &       0.149** &       0.030** \\
                    &      (0.02)   &      (0.04)   &      (0.05)   &      (0.07)   &      (0.01)   \\
log(TA Length)    &      -0.741***&      -0.593***&      -0.981***&      -1.460***&      -0.285***\\
                    &      (0.11)   &      (0.13)   &      (0.19)   &      (0.26)   &      (0.05)   \\
Constant            &       4.326***&       2.779***&       4.330***&       5.697***&       3.246***\\
                    &      (0.68)   &      (0.87)   &      (1.15)   &      (1.64)   &      (0.43)   \\
                    \midrule
 \multicolumn{6}{c}{\textbf{\emph{Average Marginal Effect of Size}}} \\
 $\frac{\partial\:\mathbbm{1}_{Follow\:up}}{\partial\:log(Size)}$ & 0.020** & 0.036*** & 0.020*  & 0.031** & 0.030**\\
  & (0.01) & (0.01) &  (0.01) & (0.01) & (0.01) \\
 \midrule
 \multicolumn{6}{c}{\emph{Fixed Effects}} \\
 \midrule
 Publisher & - & \checkmark & \checkmark & \checkmark & \checkmark \\
 Start year & - & - & \checkmark & \checkmark & \checkmark\\
Country & - & - & - & \checkmark & \checkmark\\
  \midrule    
Pseudo R$^2$         &       0.050   &       0.084   &       0.316   &       0.442   &               \\
R$^2$               &               &               &               &               &       0.528   \\
BIC                 &     925.992   &     839.784   &     619.287   &     613.587   &     725.022   \\
N                   &         752   &         716   &         643   &         560   &         752   \\
\bottomrule
    \end{tabular}\\
    {\footnotesize $* p<0.10, ^{**} p<0.05, ^{***} p<0.01$}
        \end{center}
    \caption{Regression results for the relationship between a TA getting a renewal and its size and length. Estimation method: Probit/OLS following eq.~(\ref{eq3}). Dependent variable: $\mathbbm{1}_{Renew}$. AME computed following eq.~(\ref{eq4}). 95\% standard errors in brackets below, heteroskedasticity-robust and clustered on the publisher level. Results for the samples in which we exclude TAs $<10$ and TAs $<100$ annual publications are shown in Table \ref{tab:renewal2} in the appendix.}
    \label{tab:renewal1}
\end{table}

However, for probit regressions (in)significance of the coefficient does not necessarily imply (in)significance of the marginal effect. Therefore, we compute additionally the average marginal effect (AME): It captures the mean of marginal effects of increasing the size of a transformative agreement by one unit. Hence, we compute for every TA to the extent to which a marginal increase in size would increase the probability that this TA will be renewed. Since we log-transformed the size variable, we can express that in percentages. Hence, the AME captures by how much percentage points the likelihood for a renewal increases when the size of the initial contract is 1\% larger. This is expressed as follows: 
\begin{equation}
    AME(Size) = \frac{1}{N}\sum\frac{\partial E[\mathbbm{1}_{Renew}|\sigma_i,\delta_i, \boldsymbol{\theta}]}{\partial \sigma_i}
    \label{eq4}
\end{equation}
$N$ means the number of \emph{first} TAs, i.e, contracts that have not been renewed (yet). This reduces the sample as all follow-up contracts are excluded. Furthermore, $\boldsymbol{\theta}$ is the vector of fixed effects, which varies across regressions. The \checkmark\: sign in Table \ref{tab:renewal1} indicates which actual effects $\boldsymbol{\theta}$ embodies. $\partial$ represents the incremental changes in numerator and denominator.\\
\FloatBarrier

\begin{figure}[htbp]
    \centering
    \includegraphics[width=.49\linewidth]{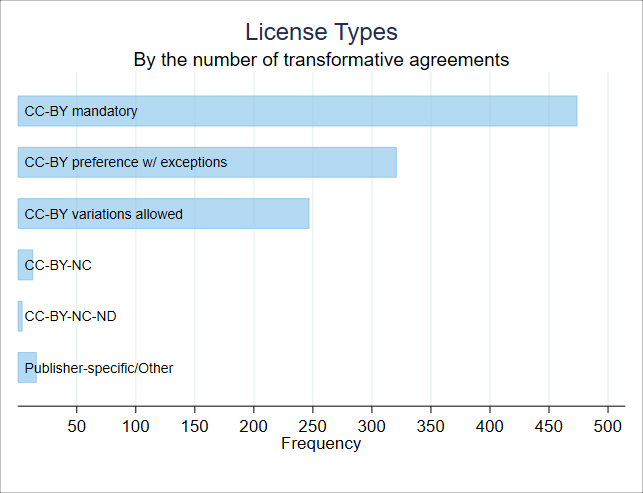}
    \includegraphics[width=.49\linewidth]{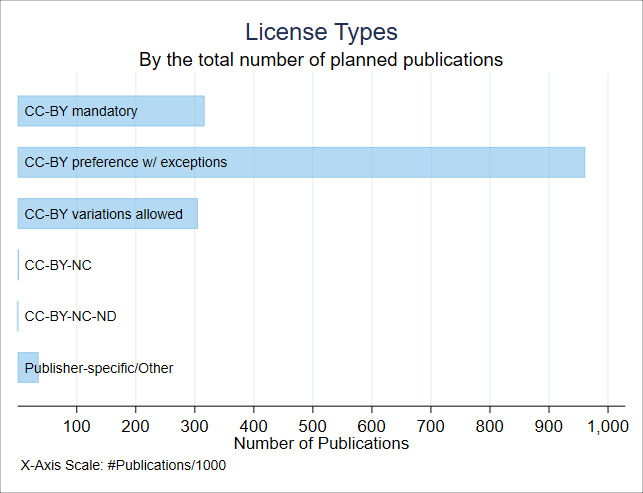}
    \caption{Aggregate Number and Total Volume of TAs by License Type, n = 1075}
    \label{fig:license_vol1}
\end{figure}

\noindent \textbf{License Types:} Several OA licensing options exist that vary in different dimensions. However, TAs are very similar in this dimension. Almost all TAs target the most flexible CC-BY license, as Figure \ref{fig:license_vol1} highlights. However, only approximately one-fourth of the contracts make this license type mandatory. Most contracts allow for exceptions, which might be due to exemptions for specific journals researchers assign with a high emphasis, but the respective publishers are unwilling to change the license model for these journals. This applies especially to the volume of covered publications. The right panel of Figure \ref{fig:license_vol1} displays the license types separated by covered volume (taking also into account the duration of the contract). It becomes obvious that exceptions are ubiquitous in TAs. Thus, the attempt to enforce strict CC-BY mandates has not yet been successful.
\FloatBarrier

\subsection{Qualitative Analysis}
\label{ssec.qual}
\vspace{-2mm}
\noindent\textbf{Capped and uncapped agreements:} In their online form, ESAC asks the following question: `\textit{How do fixed open access publishing entitlements correlate to the anticipated output? What mechanisms for risk sharing have been agreed on in cases of exceeding or not reaching this number? Please describe:}'. The answers were collected as free text. The answer is not mandatory, and 757 consortia provided information (70.42 \%). The question specifically pertains to capped agreements, so participants most commonly indicated if the TA being registered was capped or not capped. A capped agreement indicates a maximum number of publications covered under the terms of the negotiated agreement. 32.7\% of all consortia mentioned that the agreement was not capped, 29.7\% described the capping, and 37.6\% did not answer this question or the information was not sufficient to conclude from it. There is a possibility that no answer to this question indicates an uncapped agreement corresponding to the way the question was asked. 

\begin{figure}[htbp]
    \centering
    \includegraphics[width=.6\linewidth]{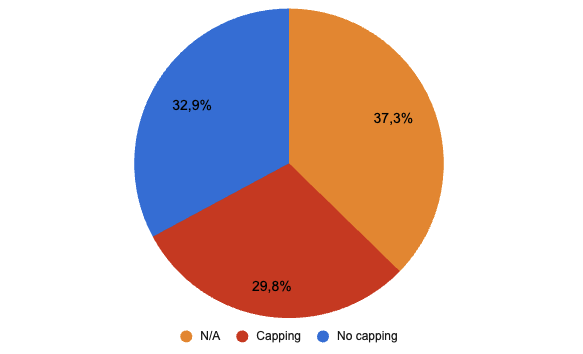}
    \caption{Share of codes indicating capping, no capping, and no answer, N=1075}
    \label{fig:capping}
\end{figure}

\noindent\textbf{Risks:} All comments were analyzed according to inductively emerging themes. Risk sharing (relevant predominantly for RAP contracts) between publishers and TA partners (consortia and libraries) refers to risks in the case of over-publishing or under-publishing according to the minimum or maximum thresholds agreed in the individual TAs. This means that equal risk sharing agrees if there are fallback options in place for libraries in case of under-publishing and, therefore, not reaching the allotted budget spent on the TA in comparison to former subscription prices. For publishers, equal risk sharing occurs if over-publishing occurs, and assumed APCs for published outputs would theoretically exceed the amount negotiated through the TA. 

In discussing risk-sharing options, the themes in Table \ref{tab:risk_codes} were identified most frequently for capped agreements. Other themes named at least once included the possibility for cap-off payments in case of reached caps, the possibility for retroactiviation of formally closed articles to OA (5), the existence of corridors for acceptable amounts for over-publishing (5), the possibility for off-setting to former subscription cost (3) and the possibility of no embargoes for Green OA publication after capping is reached (2).  

\begin{table}[htbp]
    \centering
    \footnotesize
    \begin{tabular}{p{4cm} p{6cm} c}
    \toprule
       Theme  & Description &  Frequency \\ 
       \midrule
      Roll-over possible & Coded when the possibility of roll-overs and/or transfer of credits from one year of the agreement to the next year was mentioned & 43 \\
      No roll-over possible & Coded when fixed amount of waivers or tokens only valid for the current year were mentioned & 35\\
      APC discounts &  Coded when discounted APCs after reaching caps were mentioned & 32 \\
       No refunds & Coded when no possibility of refunding in cases of under-publishing was mentioned & 23 \\
       gold OA not included & Coded when gold OA was specifically excluded or when only hybrid journals were mentioned& 22 \\
       No APC discounts & Coded when the full amount of APCs after reaching a cap was mentioned & 11 \\
       Refunds possible & Coded when a possibility of refunds in cases of under-publishing was mentioned& 6 \\
        \bottomrule
    \end{tabular}
    \caption{Frequency of occurrence of identified themes in data for capped agreements (n=317)}
    \label{tab:risk_codes}
\end{table}

Risks for libraries in conjunction with under-publishing in capped agreements were most commonly mitigated through the possibility of a roll-over period for negotiated tokens or quotas for articles. The data shows that roll-overs are mentioned most frequently. Far less often, the possibility of refunds to the library was described. In particular, possible roll-overs were mentioned more often than no roll-overs, but missing options for refunds were mentioned more often than refund possibilities. Another possible risk mitigation strategy was mentioned several times, describing the use of OA tokens retroactively to already published articles and, therefore, avoiding possibly not succeeding in reaching the minimum quota of articles for a year. 

Capping agreements inherently are a way for publishers to deal with the risk of losing income through unexpected amounts of articles without compensation through APCs. Caps are normally negotiated according to previous publication volumes, e.g. the amount of published articles in the previous year. TAs often include an increasing quota, meaning that the number of articles before reaching the cap increases each year. Many capped agreements also do not include articles in genuine gold OA journals but are focused on hybrid journals. It should be noted that uncapped agreements also mentioned the exclusion of gold OA from the TA. For libraries, if the quota is reached, APCs have to be paid for all following publications, or articles have to be published closed access. Frequently, publishers offer discounts on APCs in case of over-publishing, and there is a certain corridor for over-publishing negotiated through the TA. 
For uncapped agreements, the exclusion of Gold OA publishing was by far the most mentioned risk. Generally, uncapped agreements were considered far less risky or not risky at all. However, in some rarer cases, risk was described as equally shared, because even though OA publishing was included as unlimited (no risk of over-publishing), no refunds were possible. \\

\noindent\textbf{Workflow assessment:} ESAC asks participants to rate the quality of workflows. The questions reads: `\textit{How would you assess the practical implementation of the agreement? Please also consider the questions formulated in our workflow assessment.}' This question in the form is a non-required multiple choice question, with the answer options 'OK', 'Deficient' and 'Too early to assess'. It was answered by 666 participants (61.95\%). 
We analysed the answers by groups devided into capped and uncapped agreements. Among 349 uncapped agreements, workflows of {\color{OliveGreen}128} TAs were evaluated as `Ok,' only {\color{BrickRed}17} as `deficient,' and {\color{NavyBlue}204} as 'too early to assess.' Out of 317 capped agreements, workflows of {\color{OliveGreen}141} as `Ok,' {\color{BrickRed}26} as `deficient,' and {\color{NavyBlue}150} as `too early to assess.' The numbers are displayed in Table \ref{tab:overall_workflow}.

\begin{table}[htbp]
    \centering
    \footnotesize
    \begin{tabular}{l c r}
    \toprule
    \multicolumn{3}{c}{\textbf{\textit{Workflow Assessment}}} \\
    \midrule
       \multicolumn{3}{c}{\textit{Uncapped }(n=349)} \\ 
      {\color{OliveGreen}OK} &  {\color{BrickRed}Deficient} & {\color{NavyBlue}Too early to assess} \\
       128 & 17 & 204 \\
       36.68 \% & 4.87 \% & 58.45 \%\\ 
       \midrule
       \midrule
       \multicolumn{3}{c}{\textit{Capped} (n=317)}  \\
 {\color{OliveGreen}OK} & {\color{BrickRed}Deficient} & {\color{NavyBlue}Too early to assess} \\
      141 &  26 & 150 \\
       44.48 \% & 8.2 \% &  47.32 \%\\
        \bottomrule
    \end{tabular}
    \caption{Workflow Assessment of uncapped and capped agreements}
    \label{tab:overall_workflow}
\end{table}

Comments were possible as a free text option. There were 241 comments, which we analyzed according to commonly occurring themes. Positive comments varied, but a good automated workflow with CCC RightsLink and the existence of a dashboard for real-time monitoring were two outstanding recurring themes. Negative comments also varied in themes, with complicated and manual workflows, as well as the lack of possibility of monitoring through the library mentioned the most. Table \ref{tab:risk_codes} shows the three most frequently occuring themes for positive and negative comments.

\begin{table}[htbp]
    \centering
    \footnotesize
    \begin{tabular}{p{5cm} p{6cm}  c}
    \toprule
       \textbf{Theme positive} & Description & Frequency  \\ \midrule
       CC RightsLink & Coded when CC Rightslink was positively mentioned &35 \\
      Dashboard & Coded when dashboards by publishers for positively mentioned &14 \\
    Automated Workflow & Coded when automated workflows were positively mentioned &12 \\ 
    \midrule
    \midrule
    \textbf{Theme negative} & Description & Frequency \\ \midrule
     Complicated & Coded when tedious or complicated workflows were negatively mentioned &19 \\
        Manual Workflow & Coded when manual workflows were negatively mentioned &15 \\
       No monitoring of workflow possible & Coded when lack of possibility to monitor workflows was negatively mentioned &15 \\
        \bottomrule
    \end{tabular}
    \caption{Frequency of the top three identified themes in comments on workflow (n=241)}
    \label{tab:risk_codes}
\end{table}

\noindent\textbf{Overall assessment:} The consortia can comment on their “Overall Assessment” via the ESAC form when registering their agreements. The question reads as follows: \textit{`What is the agreement's impact on the overall open access transition? Do you think it is sustainable with regard to costs and operations? Which elements and mechanisms would have to be adjusted in future agreements? Please share your overall assessment of this agreement with the community!'}. We coded comments according to their general sentiments as {\color{OliveGreen} `generally positive,'} {\color{BrickRed}`generally negative,'} and {\color{NavyBlue}`too early to assess.'} In total, 422 forms contained comments on the overall assessment. {\color{OliveGreen}238} comments were coded as `generally positive,' {\color{BrickRed}126} as `generally negative,' and {\color{NavyBlue}58} as `too early to assess'. 653 forms did not contain comments. 

\begin{table}[htbp]
    \centering
    \footnotesize
    \begin{tabular}{l c r}
    \toprule
    \multicolumn{3}{c}{\textbf{\textit{Overall Assessment}}} \\
    \midrule
      {\color{OliveGreen}Generally positive } &  {\color{BrickRed}Generally negative} & {\color{NavyBlue}Too early to assess} \\
       238 & 126 & 58 \\
        \bottomrule
    \end{tabular}
    \caption{Overall assessment of TAs in comments (n=422)}
    \label{tab:overall_general}
\end{table}

For uncapped agreements, the overall assessment was mostly generally positive {\color{OliveGreen} (111)}, with some generally negative {\color{BrickRed}(49)} and a few `too early to assess' {\color{NavyBlue}(15)}. For 174 out of 349 uncapped TAs, there was no comment on overall assessment. Out of the 317 capped agreements, {\color{OliveGreen}80} were evaluated as `generally positive,' {\color{BrickRed}56} as `generally negative,' and {\color{NavyBlue}20} as `too early to assess.' For 161 capped agreements, no comments on the overall assessment exist, as one can draw from Table \ref{tab:overall_overall}, which provides all numbers.

\begin{table}[htbp]
    \centering
    \footnotesize
    \begin{tabular}{l c c r}
    \toprule
    \multicolumn{4}{c}{\textbf{\textit{Overall Assessment: Capped/Uncapped Agreements}}} \\
    \midrule
       \multicolumn{4}{c}{Uncapped (n=349)} \\
       {\color{OliveGreen} Positive} & {\color{BrickRed}Negative} & {\color{NavyBlue}Too early to assess} & N/A \\
       111 & 49 & 15 & 174 \\
       31.81 \% & 14.04 \% & 4.3 \% & 49.86 \%\\ 
       \midrule
       \midrule
       \multicolumn{4}{c}{Capped (n=317)} \\
       {\color{OliveGreen} Positive} & {\color{BrickRed}Negative} & {\color{NavyBlue}Too early to assess} & N/A \\
       56 & 80 & 20 & 161 \\
       17.67\% & 25.24 \% & 6.31 \% & 50.79 \%\\
        \bottomrule
    \end{tabular}
    \caption{Overall Assessment of uncapped and capped agreements}
    \label{tab:overall_overall}
\end{table}

The difference between capped and uncapped agreements in the workflow assessment was noticeable, with capped agreements generally receiving more `deficient' scoring than uncapped agreements. Even more significant is the difference between capped and uncapped agreements with respect to the overall assessment: While uncapped agreements received far more positive assessments than negative assessments, capped agreements tended to be generally more negative than positive. \\

\noindent\textbf{Negative comments:} We evaluated {\color{BrickRed}negative valence} comments and labeled them with subcodes. These subcodes, in turn, can be clustered into specific themes, which are listed in Table \ref{tab:neg_codes}. Of the total of 1.075 agreements examined, we found 126 with a critical statement in the `Overall Assessment' field. 

\begin{table}[htbp]
    \centering
    \footnotesize
    \begin{tabular}{p{5cm} p{6cm}  c}
    \toprule
       \textbf{\textit{Theme}} & \textbf{\textit{Description}} & \textbf{\textit{Frequency}}  \\ \midrule
       Contract terms & Coded when contract terms were negatively mentioned & 87 \\
       OA exploitation & Coded when under-publishing or lack of OA increase was mentioned & 22 \\
       Financial issues & Coded when issues with budgeting and affordability were mentioned & 18 \\
       Publishing services & Coded when issues with services by publishers were mentioned & 18 \\
       Author engagement & Coded when lack of author acceptance or awareness were mentioned & 5 \\
        \bottomrule
    \end{tabular}
    \caption{Frequency of identified themes in negative comments on TAs (n=126), multi-codes included.}
    \label{tab:neg_codes}
\end{table}

Thematically, critical comments on contract terms (87 TAs),  OA exploitation (22), publishing services (18), and financial issues (18) dominate ahead of author engagement (5) - although it should be noted that the responses come from libraries, not from authors themselves. We also looked into these higher-level themes in more depths in order to better understand the nature of these comments.

As far as the contract terms were concerned, capping was criticized in 15 cases and the coverage of the agreement in 37 cases (including the lack of gold OA journals in 25 cases and the lack of closed access content in 15 cases; three consortia criticized the lack of both types of content). One consortium criticized the lack of an option to carry over unused vouchers to the next year (`no roll-over'), while two agreements (with Elsevier and Wiley) were criticized due to the lack of Plan S compliance. 21 agreements received negative comments because article fees are factored based on former subscription expenses, so the underlying cost calculation does not comply with the PAR principles and, therefore, does not meet the requirements of a transition to OA. For eight agreements, the OA costs were critically assessed; in four cases, they were due to pricing (the unexplained fixation of publication costs) and in four cases, they were due to the structure of pricing (unbalanced cost structure). 

In terms of OA exploitation, for two agreements, it was noted that the available OA publication volume was not fully utilized. For ten agreements, the consortia were concerned that the level of OA publication was insufficient to flip/transform the journals to OA. For another ten agreements, low OA uptake was noted. 

Regarding financial issues, budgeting problems were mentioned as critical for four agreements. Skepticism was reported on sustainability regarding affordability for 14 agreements. Five comments related to author engagement: Four mentioned a low author awareness, one a lack of author acceptance. This was the case for the TA between Taylor \& Francis and the Portland State University Library.

Concerning publishing services, there were critical comments regarding reporting mechanisms by publishers (three cases), workflows (15 cases), and, in one case, the potentially negative effects of the APC model on the quality of scientific publications. \\

\noindent\textbf{Positive comments:} 238 comments were coded as {\color{OliveGreen} generally positive}. We also used inductive coding for this subset of comments and looked for common themes. There was some variation in the comments, but the five most frequently mentioned arguments for evaluating TAs as generally positive emerged. These are shown in Table \ref{tab:positive}. 

\begin{table}[htbp]
    \centering
    \footnotesize
    \begin{tabular}{p{5cm} p{6cm}  c}
    \toprule
       \textbf{\textit{Theme }} & \textbf{\textit{Description}} & \textbf{\textit{Frequency}}  \\ \midrule
      Transition and transformation & Coded when transition to full OA and/or transformation of journals was mentioned &101\\
      Sustainable regarding costs & Coded when costs were seen as sustainable & 66 \\
      No cap & Coded when unlimited reading and publishing was mentioned & 58 \\
      Good workflows & Coded when workflows were praised as being less complicated and well done by the publisher & 29 \\
      Transparency & Coded when prices were described as transparent & 15 \\
        \bottomrule
    \end{tabular}
    \caption{Frequency of identified themes in positive comments on TA (n=238), multi-codes included.}
    \label{tab:positive}
\end{table}

Most often, TAs were considered to be a good step toward the transition to full OA and the transformation of whole journals. Comments included sentences such as `\textit{This agreement is a step forward toward 100\% Open Access in (country)}', `\textit{This agreement transitions 100\% of (country)'s research publishing output towards open access}', and `\textit{The Agreement, providing unlimited OA publications both in hybrid and gold journals, offers 100\% coverage of the articles of corresponding authors affiliated with the subscribing institutions. The transition is therefore completed}'. 

The question specifically asked participants to comment on the sustainability of workflows and the costs of a TA. Sustainability regarding costs was mentioned 66 times, also referring to cost neutrality or a minimal increase of costs in line with other benefits, such as improved workflows for researchers and contractors (libraries). Sustainability also referred to adequate capping, if capped agreements were regarded as positive.

Unlimited OA publishing (no cap) was another frequent comment for positively rated agreements. This concurs with results in table \ref{tab:overall_overall}, showing that uncapped agreements are generally rated more likely as positive than capped agreements. 

Good workflows, which suppress complicated communications between libraries and publishers and make the handling of publications published under a TA as easy as possible for libraries, also contributed to positive comments. This also included praise for workflows around monitoring.

In addition, transparency, both in terms of pricing models and workflows and reporting from the publisher, was seen as overall positive. Here, TAs help to reduce intransparent conditions when dealing with APCs for different publishers and have a positive effect on clarity about library publishing progress.

\section{Discussion}
\label{sec.disc}

Research managers, administrators, and librarians are not economists. And this is a good thing. However, their enthusiasm for tearing down paywalls in the digital age and opening cutting-edge research to everyone maneuvered research institutions into a situation in which the acclaimed transformation is halfway stuck. TAs - meant as transitory bridges- delivered in the domain of fostering OA: Plenty of papers that would otherwise have been published in subscription journals are now freely available because of the OA component that became contractual in most agreements.

However, the leading publishers are obviously a vital part of this transition. This is due to the lagged reputation mechanisms (\cite{schmal_x_2023}), which endow long-standing journals with substantial market power. These are often owned by large publishers. Thus, the major commercial actors are, for better or worse, an integral part of every transition - they even seem to be able to use this to their advantage \cite{shu_oligopoly_2024}. As our results show, Elsevier, Springer Nature, and Wiley, in particular, manage to translate their market position into substantially longer and larger contracts. While the latter is simple, given the enormous amount of covered journals, the contract length is a non-negligible accounting advantage as it ensures longer-running cash flows and blocks money from the libraries and research institutions that cannot be spent on other publishers or projects. However, negotiations are often difficult and lengthy, and making longer contracts reduces the number of times new negotiations have to be initiated for both contract partners, which may be preferable by both parties. 

Interestingly, western countries, Germany, the Netherlands, and the United Kingdom, negotiate much longer contracts with the large publishers Elsevier, Springer Nature, and Wiley than TAs with other publishers. The publishers' side might not be the only one responsible for that. In addition, universities or institutional consortia may be interested in longer contracts if negotiations are resource-intensive (\cite{mittermaier_aus_2017}). However, this asymmetrically targets larger publishers as it is, all things equal, more difficult to negotiate a contract for a larger and potentially more diverse set of journals and outlets. Hence, universities may inadvertently strengthen large publishing houses simply because they want to avoid renegotiation that is too frequent. This advantage stems from sheer size, which provides larger publishers, but mostly the `big 3,' with a longer planning horizon, better cash flow planning, and a longer time without the need to deploy staff to negotiate contract renewals. These are significant competitive advantages for these publishers, and the audience needs to be aware of these strategic implications. It is particularly notable because especially representatives of the research administration in these three countries often call for systemic changes to academic publishing that slow down commodification, tackle the power of the large publishers, and strengthen non-profit publishing initiatives.

The qualitative analysis of the overall assessment listed in the ESAC database emphasizes that several times, gold OA journals were not covered by TAs, even though the demand side would have appreciated it. Unfortunately, even if the gold OA journals were part of a TA, we do not know whether these journals have benefited from TAs in the same way as the covered hybrid journals. This is particularly questionable because, although the costs of hybrid publications in TAs are usually borne by the libraries, this is often not the case for gold OA, but only discounts are granted, so the payment lies with the authors, which puts gold OA at a structural disadvantage. The exclusion of gold OA publishing from both capped and uncapped TAs raises important questions about the balance of interests between publishers and libraries in the OA transition. Gold OA, which allows immediate and unrestricted access to research outputs, is a key component of the broader OA movement. However, its exclusion from many TAs, especially in uncapped agreements, suggests a strategic move by publishers to retain revenue from APCs. 

Regarding the costs of TAs, many positive assessments mention that there is no or just a slight increase. However, such schemes reinforce the status quo in terms of pricing. The influential `Max Planck Digital Library' whitepaper by \cite{schimmer_disrupting_2015} calculated that while subscription-based publishing is associated with costs of 3,800-5,000 EUR per published paper, they draw from several reports and estimations that a full OA landscape could ensure sustainable publishing for publication fees below 2,000 EUR per paper. Because TAs are not the goal but only a step in this direction, it is reasonable that the costs are higher. However, a decreasing path should be visible in order to reach significantly lower costs in the medium term. The objection is that also large gold open access publishers such as Frontiers charge APCs well above that threshold. Frontiers, for example, charged in 2022 on average 2,270 USD per publication.\footnote{In 2022. See the `Fee Policy' report of Frontiers on their website: \url{https://www.frontiersin.org/about/fee-policy}, last checked 13 September 2024.} However, this is rather a signal that the high costs of academic publishing are not so much a question of access but of competition in the market, see also \cite{schmal_wettbewerb_2024}.

Libraries' perception of risk in TAs reflects a notable distinction between \textit{capped} and \textit{uncapped} agreements. Uncapped agreements are generally viewed as less risky or even risk-free, despite the potential for under-publishing. This perspective suggests that libraries prioritize the risks associated with over-publishing, where they might face unexpected costs if article submissions exceed anticipated numbers. In contrast, the risk of under-publishing, where the library does not fully utilize the allocated budget, seems to be of lesser concern. This could be due to libraries' reliance on historical publishing patterns or confidence in achieving expected output levels, but also due to the fact that the budget spent on TAs has already been allocated and under-publishing would not change this allocation. The relative disregard for under-publishing risks may also indicate a preference for the predictability of uncapped agreements, where the absence of limits provides a perceived security against sudden financial burdens and cumbersome negotiations on how to cover the unexpected extra costs, overshadowing the potential inefficiencies in budget utilization. 

Libraries, while initially enthusiastic about TAs as a means to foster OA, now find themselves in a complex situation where these agreements, intended as temporary solutions, have become long-term commitments that are challenging to exit. At the same time, they find themselves in the inconvenient situation of being both volume consumers (reading component) and - in the case of the PAR model - managers of unpredictable costs whose origin is beyond their control. The strategic advantage gained by large publishers through the negotiation of extended contracts effectively traps libraries in a cycle of dependency. By locking in these agreements for longer periods, libraries inadvertently restrict their own flexibility, both financially and operationally. This situation is intensified by the fact that these contracts often do not cover gold OA journals, thereby limiting the potential for a full transition to OA and keeping libraries tethered to a system that perpetuates the status quo rather than disrupts it. 

Moreover, the apparent security of uncapped agreements, which are perceived as less risky by libraries due to their focus on mitigating over-publishing costs, has its own set of drawbacks. By focusing predominantly on the risks of over-publishing, libraries may overlook the inefficiencies and potential losses associated with TAs in general, specifically for RAP agreements. This selective risk management approach not only reinforces the publishers' stronghold on the OA transition but also limits the libraries' ability to reallocate resources towards more innovative or cost-effective publishing models. As a result, rather than serving as a bridge to a more open and equitable scholarly communication landscape, TAs may have inadvertently solidified the dominance of a few large publishers, trapping libraries in agreements that may be financially unsustainable and operationally restrictive in the long term.
 
\subsection{Limitations}
\label{ssec.lim}
It is essential to acknowledge limitations to our analysis, especially those imposed by the data used in this analysis. Exploiting not only the aggregated dataset provided as a spreadsheet but also the detailed contract-specific metadata is a unique source of information that remained unused up to now. Nevertheless, the data comes with some caveats: It was collected in a non-standardized way, using a voluntary form on the ESAC website with quite a large amount of free text answer options. The heterogeneous quality of the data led to challenges in aggregating and comparing information between different agreements and publishers. This was especially true for the qualitative analysis since the data suffered from incomplete, very short answers or answers that were copied/pasted for each contract, leading to many duplicates. 

Our reliance on secondary data sources, which we reused for this analysis, represents the best information available at the time of the study. However, this approach inherently carries the limitation of relying on data that may not capture the most current developments or nuanced contractual terms due to delays in reporting or confidentiality clauses. The lack of a standardized and centralized database for TAs exacerbates this problem, making it difficult to verify the completeness and accuracy of the information gathered. This limitation underscores the challenges science studies researchers and information professionals face when attempting to assess the rapidly evolving landscape of open access publishing using publicly available data.

These data limitations may have affected our findings, potentially underrepresenting certain trends or overemphasizing others. For instance, incomplete data on the coverage of gold OA journals within TAs may have led to an underestimation of their exclusion's impact on the overall open access movement. To enhance the validity and reliability of future research, there is a pressing need for more standardized and transparent data collection practices. Establishing uniform reporting standards and encouraging greater openness from publishers and institutions would facilitate more detailed and accurate analyses. Better data quality would not only strengthen the conclusions drawn in this paper but also support stakeholders in making more informed decisions during the ongoing transition to open access.

Thus, as we join the chorus of those calling for better data for research on the transformation of academic publishing, we also push back on the critique that research working with the current set of data may be invalid. It is, of course, true that any quantitative analysis of such complex contracts must truncate some information, and any qualitative analysis depends on the quality of the voluntarily free-text information. Nevertheless, independent research is desperately needed in light of the many involved stakeholders with their own interests, the high amounts of money spent on reading and publishing in plenty of ways and the dual role publishers play in this situation. Not conducting an analysis like ours but waiting for `better' data would imply a blind flight. Quite the opposite, fishing in murky waters does not only provide \emph{some} evidence, it may also initiate additional and advanced data collection. We consider our analysis important, not even though there exist limitations to the data quality, but because of their existence. 

\section{Conclusion}
\label{sec.conc}

The present paper offers a comprehensive analysis of virtually all TAs that began in 2013 and continue to this day. Despite the original intent to serve as `transitional bridges' towards a publishing system consisting solely of fully OA journals, evidence suggests that these agreements are not achieving their intended outcomes at the expected pace or even at all.

Our combined qualitative and quantitative approach, based on the extensive contract characteristics listed in the ESAC database, highlights that rather than acting as a bridge to a fully OA environment, TAs may instead be reinforcing the hybrid model. Here, journals continue to offer both OA and subscription-based options. It raises significant concerns about the sustainability and effectiveness of TAs to transform the entire publishing market. The extended duration and increased volume of these agreements, particularly with major publishers, suggest that institutions may inadvertently perpetuate the status quo rather than promote changes in the market characteristics that could lead to increased competition. The concentration of agreements with a few dominant players, combined with the tendency that initially larger contracts are more likely to receive a renewal, implies that larger publishers may gain a competitive advantage. This could potentially stifle diversity and innovation in the academic publishing landscape, a concerning prospect for the future.

The financial implications are also significant. The data from 1,075 analyzed contracts suggest that the long-term financial burden on libraries and research institutions may not permanently decrease as they may become locked into ongoing negotiations and payments without realizing substantial reductions in subscription costs. The apparent failure of many journals to fully transition to OA under these agreements further exacerbates this issue, leading to concerns about the long-term sustainability of the current TA model. \\ 

\textit{Are research institutions `trapped' in TAs?} In light of our analysis, the answer, unfortunately, seems to be `yes.' Therefore, it's crucial for all academic stakeholders to critically evaluate the role of TAs in the broader OA movement and in their particular institutional environment. TAs have undoubtedly promoted the adoption of OA, but their current path may not align with the overarching goals of `democratizing' access to knowledge and reducing the financial burden on the academic community. Future efforts should focus more on ensuring that TAs genuinely contribute to the transition of journals to OA and do not merely serve as a temporary measure that, up to now, seem to perpetuate existing power structures and financial dependencies in academic publishing.

\pagebreak
\begin{singlespace}
\printbibliography
\addcontentsline{toc}{section}{\protect\numberline{}References}
\end{singlespace}

\newpage
\begin{singlespace}
\subsection*{Acknowledgments}
We are grateful for valuable feedback from Coleen Campbell, Sascha Lauer, Bernhard Mittermaier, and helpful research support and feedback from Dorothea Strecker. Their constructive feedback significantly enhanced the quality of this work. All errors and opinions are solely those of the authors. Acknowledging feedback from commentators does not imply that they endorse the content of this paper.

\subsection*{Author contributions}
W.~Benedikt Schmal, Laura Rothfritz, Ulrich Herb -- Conceptualization, Data curation, Formal analysis, Investigation, Methodology, Validation, Writing - original draft, Writing - review \& editing \\
W. Benedikt Schmal, Laura Rothfritz, Dorothea Strecker -- Visualization

\subsection*{Competing Interests}
The authors declare they do not have any competing interests.

\subsection*{Data Availability Statement}
Replication code, scripts, and data used for this study are publicly available on Zenodo: \url{https://zenodo.org/doi/10.5281/zenodo.13453629}.
\end{singlespace}

\newpage
\section*{Appendix}
\label{sec.app}
\renewcommand{\thefigure}{F\arabic{figure}}
\setcounter{figure}{0}
\renewcommand{\thetable}{A\arabic{table}}
\setcounter{table}{0}

\begin{figure}[htbp]
    \centering
    \includegraphics[width=0.6\linewidth]{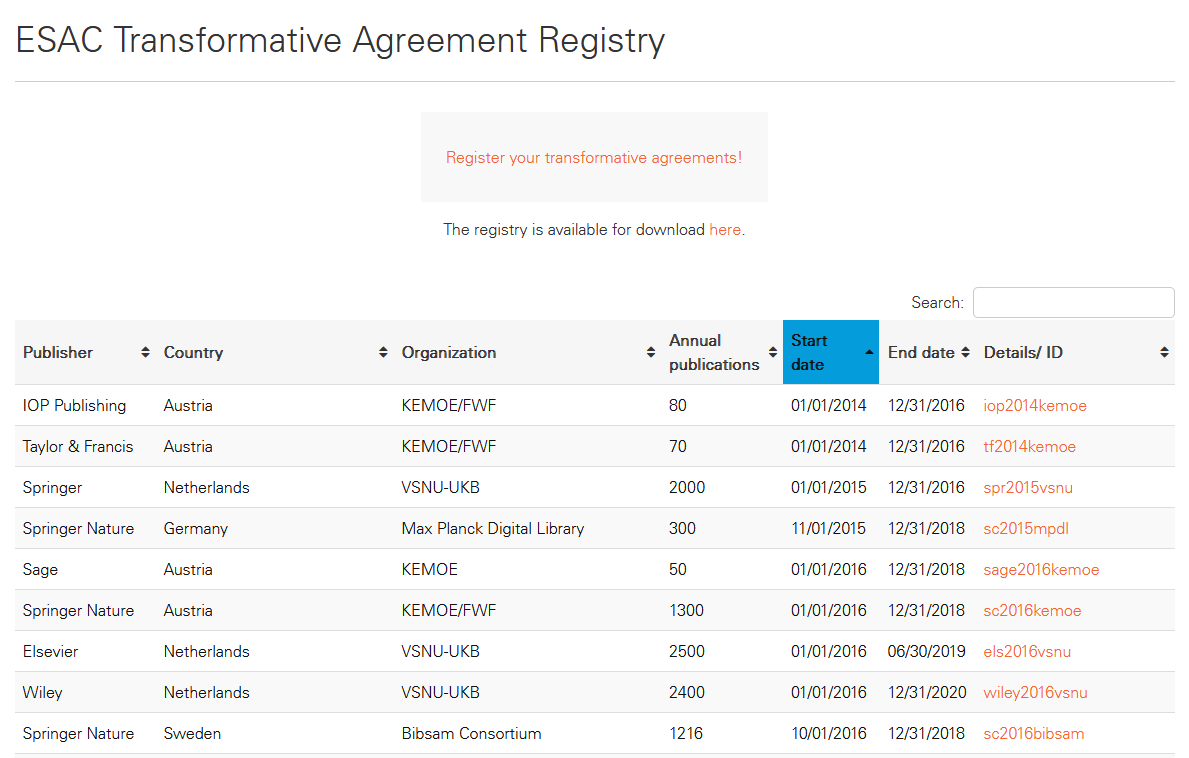}
    \includegraphics[width=0.6\linewidth]{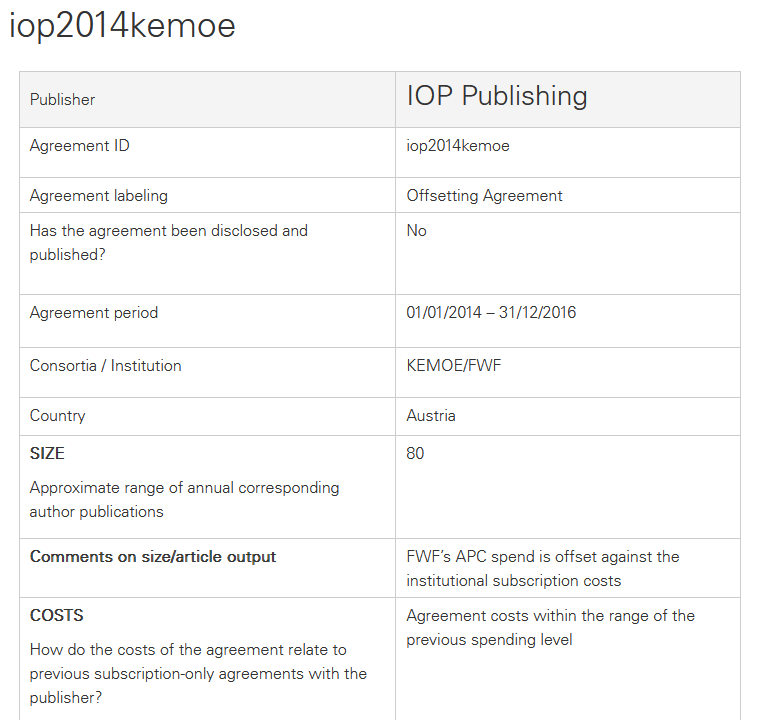}
    \caption{Screenshot of the contents of the ESAC database. Upper screenshot from \url{https://esac-initiative.org/about/transformative-agreements/agreement-registry/}, lower screenshot from \url{https://esac-initiative.org/about/transformative-agreements/agreement-registry/iop2014kemoe/}. Website accessed on 12 August 2024.}
    \label{fig:esac_screenshots}
\end{figure}

\begin{table}[htbp]
\centering
\scriptsize
\begin{tabular}{lrrr}
\toprule
\textbf{Publisher}  & \textbf{Frequency} & \textbf{Share} & \textbf{Cumulative} \\
\midrule
Cambridge UP                            & 75    & 6.98    & 6.98   \\
IOP Publishing                          & 59    & 5.49    & 12.47  \\
Oxford UP                               & 59    & 5.49    & 17.95  \\
Royal Soc Chem                          & 53    & 4.93    & 22.88  \\
Springer Nature                         & 53    & 4.93    & 27.81  \\
Taylor \& Francis                       & 51    & 4.74    & 32.56  \\
Wiley                                   & 51    & 4.74    & 37.30  \\
ACS                                     & 46    & 4.28    & 41.58  \\
Company of Biologists                   & 44    & 4.09    & 45.67  \\
ACM                                     & 42    & 3.91    & 49.58  \\
The Royal Society                       & 42    & 3.91    & 53.49  \\
\hline
Elsevier                                & 41    & 3.81    & 57.30  \\
SAGE                                    & 35    & 3.26    & 60.56  \\
John Benjamins Publishing               & 32    & 2.98    & 63.53  \\
Walter de Gruyter                       & 30    & 2.79    & 66.33  \\
AIP Publishing                          & 28    & 2.60    & 68.93  \\
Microbiology Society                    & 27    & 2.51    & 71.44  \\
Portland Press                          & 27    & 2.51    & 73.95  \\
IEEE                                    & 23    & 2.14    & 76.09  \\
The Company of Biologists               & 23    & 2.14    & 78.23  \\
Rockefeller UP                          & 21    & 1.95    & 80.19  \\
Emerald                                 & 19    & 1.77    & 81.95  \\
Karger                                  & 16    & 1.49    & 83.44  \\
SPIE                                    & 16    & 1.49    & 84.93  \\
BMJ                                     & 15    & 1.40    & 86.33  \\
Brill                                   & 14    & 1.30    & 87.63  \\
Wolters Kluwer Health                   & 14    & 1.30    & 88.93  \\
Geological Society London               & 10    & 0.93    & 89.86  \\
Cold Spring Harbor Laboratory Press & 9     & 0.84    & 90.70  \\
\hline
IWA Publishing                          & 7     & 0.65\%    & 91.35\%  \\
Thieme                                  & 7     & 0.65\%    & 92.00\%  \\
American Physical Society               & 6     & 0.56\%    & 92.56\%  \\
European Respiratory Society            & 6     & 0.56\%    & 93.12\%  \\
The Electrochemical Society             & 5     & 0.47\%    & 93.58\%  \\
Akadémiai Kiadó                         & 4     & 0.37\%    & 93.95\%  \\
Bioscientifica                          & 4     & 0.37\%    & 94.33\%  \\
EDP Sciences                            & 4     & 0.37\%    & 94.70\%  \\
Radiological Society of North America   & 4     & 0.37\%    & 95.07\%  \\
Royal Irish Academy                     & 4     & 0.37\%    & 95.44\%  \\
American Physiological Society          & 3     & 0.28\%    & 95.72\%  \\
American Psychological Association      & 3     & 0.28\%    & 96.00\%  \\
Hogrefe                                 & 3     & 0.28\%    & 96.28\%  \\
IOS Press                               & 3     & 0.28\%    & 96.56\%  \\
National Academy of Sciences            & 3     & 0.28\%    & 96.84\%  \\
The Electrochemical Society (ECS)       & 3     & 0.28\%    & 97.12\%  \\
Bentham Science Publishers              & 2     & 0.19\%    & 97.30\%  \\
Bristol University Press                & 2     & 0.19\%    & 97.49\%  \\
CSIRO                                   & 2     & 0.19\%    & 97.67\%  \\
Canadian Science Publishing             & 2     & 0.19\%    & 97.86\%  \\
Cogitatio Press                         & 2     & 0.19\%    & 98.05\%  \\
MA Healthcare                           & 2     & 0.19\%    & 98.23\%  \\
Royal College of General Practitioners  & 2     & 0.19\%    & 98.42\%  \\
S. Karger AG                            & 2     & 0.19\%    & 98.60\%  \\
Society for Neuroscience                & 2     & 0.19\%    & 98.79\%  \\
Trans Tech                              & 2     & 0.19\%    & 98.98\%  \\
World Scientific                        & 2     & 0.19\%    & 99.16\%  \\
American Meteorological Society         & 1     & 0.09\%    & 99.26\%  \\
American Society of Mechanical Engine.. & 1     & 0.09\%    & 99.35\%  \\
EDP Science                             & 1     & 0.09\%    & 99.44\%  \\
Edward Elgar Publishing Ltd             & 1     & 0.09\%    & 99.53\%  \\
IGI Global                              & 1     & 0.09\%    & 99.63\%  \\
Inter-Research Science Publisher        & 1     & 0.09\%    & 99.72\%  \\
Mary Ann Liebert, Inc.                  & 1     & 0.09\%    & 99.81\%  \\
Optica                                  & 1     & 0.09\%    & 99.91\%  \\
White Horse Press                       & 1     & 0.09\%    & 100\% \\
\midrule
Total                                   & 1,075 & 100\%  &   \\
\bottomrule 
\end{tabular}
\caption{Full list of publishers and their number of TAs}
\label{tab.publ_all}
\end{table}

\begin{table}[htbp]
\centering
\scriptsize
\begin{tabular}{lrrr}
\toprule
\textbf{Publisher}  & \textbf{Frequency} & \textbf{Share} & \textbf{Cumulative} \\
\midrule
Germany        & 126      & 11.72\%                 & 11.72\%  \\
Netherlands    & 118      & 10.98\%                 & 22.70\% \\
United Kingdom & 96       & 8.93\%                  & 31.63\%  \\
United States  & 83       & 7.72\%                  & 39.35\%  \\
Austria        & 63       & 5.86\%                  & 45.21\%  \\
Hungary        & 49       & 4.56\%                  & 49.77\%  \\
Sweden         & 42       & 3.91\%                  & 53.67\%  \\
\hline
Ireland        & 41       & 3.81\%                  & 57.49\%  \\
Australia      & 40       & 3.72\%                  & 61.21\%  \\
Finland        & 39       & 3.63\%                  & 64.84\%  \\
Switzerland    & 38       & 3.53\%                  & 68.37\%  \\
Norway         & 35       & 3.26\%                 & 71.63\%  \\
Spain          & 27       & 2.51\%                  & 74.14\%  \\
Saudi Arabia   & 25       & 2.33\%                  & 76.47\%  \\
Slovenia       & 25       & 2.33\%                  & 78.79\%  \\
Czech Republic & 21       & 1.95\%                  & 80.74\%  \\
Israel         & 19       & 1.77\%                 & 82.51\%  \\
Italy          & 19       & 1.77\%                  & 84.28\%  \\
Hong Kong      & 18       & 1.67\%                  & 85.95\%  \\
South Africa   & 14       & 1.30\%                  & 87.26\%  \\
Canada         & 13       & 1.21\%                  & 88.47\%  \\
Poland         & 13       & 1.21\%                  & 89.67\%  \\
Portugal       & 12       & 1.12\%                  & 90.79\%  \\
\hline
Greece         & 11       & 1.02\%                  & 91.81\%  \\
Japan          & 10       & 0.93\%                  & 92.74\%  \\
Denmark        & 8        & 0.74\%                  & 93.49\%  \\
Turkey         & 7        & 0.65\%                  & 94.14\%  \\
Belgium        & 6        & 0.56\%                  & 94.70\%  \\
France         & 6        & 0.56\%                  & 95.26\%  \\
China          & 5        & 0.47\%                  & 95.72\%  \\
Colombia       & 5        & 0.47\%                  & 96.19\%  \\
Latvia         & 5        & 0.47\%                  & 96.65\%  \\
Lithuania      & 5        & 0.47\%                  & 97.12\%  \\
South Korea    & 5        & 0.47\%                  & 97.58\%  \\
Qatar          & 3        & 0.28\%                  & 97.86\%  \\
Slovakia       & 3        & 0.28\%                  & 98.14\%  \\
Luxembourg     & 2        & 0.19\%                  & 98.33\%  \\
Serbia         & 2        & 0.19\%                  & 98.51\%  \\
Albania, Armenia, Azerbaijan, Belarus.. & 1 & 0.09\% &  98.60\%   \\
Albania, Armenia, Azerbaijan, Belarus.. & 1 & 0.09\% &  98.70\%    \\
Albania, Armenia, Azerbaijan, Botswan.. & 1 & 0.09\% &  98.79\%    \\
Albania, Armenia, Azerbaijan, Congo (.. & 1 & 0.09\% &  98.88\%     \\
Albania, Armenia, Azerbaijan, Congo, .. & 1 & 0.09\% &  98.98\%    \\
Albania, Armenia, Azerbaijan, Congo, .. & 1 & 0.09\% &  99.07\%     \\
Albania, Armenia, Azerbaijan, Congo, .. & 1 & 0.09\% &  99.16\%     \\
Croatia        & 1        & 0.09\%                  & 99.26\%  \\
Estonia        & 1        & 0.09\%                  & 99.35\%  \\
Ghana          & 1        & 0.09\%                  & 99.44\%  \\
Iceland        & 1        & 0.09\%                  & 99.53\%  \\
India          & 1        & 0.09\%                  & 99.63\%  \\
Moldova        & 1        & 0.09\%                  & 99.72\%  \\
Palestine      & 1        & 0.09\%                  & 99.81\%  \\
Singapore      & 1        & 0.09\%                  & 99.91\%  \\
Taiwan         & 1        & 0.09\%                  & 100\% \\
\midrule
Total & 1,075 & 100\% \\
 \bottomrule
\end{tabular}
\caption{List of countries that concluded TAs}
\label{tab.coun_all}
\end{table}

\begin{figure}[htbp]
    \centering
    \includegraphics[width=.49\linewidth]{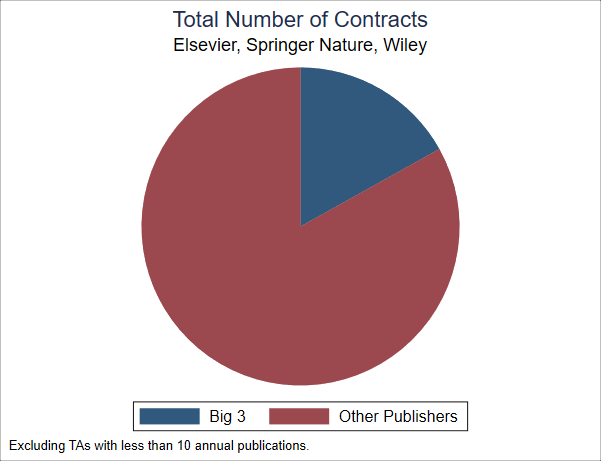}
    \includegraphics    [width=.49\linewidth]{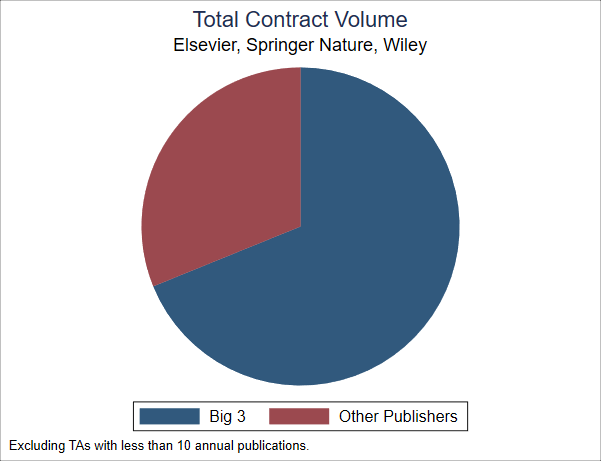}\\
    \includegraphics[width=.49\linewidth]{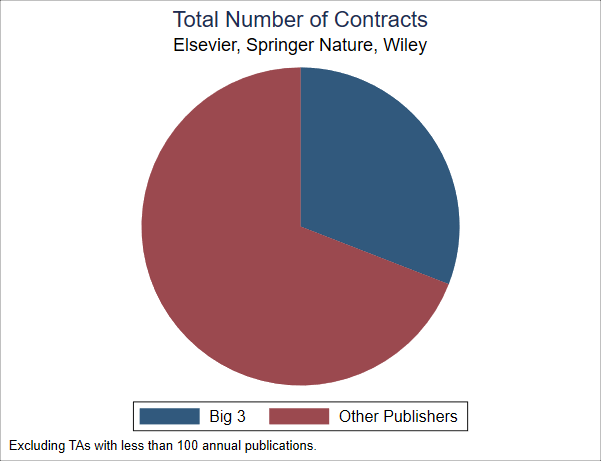}
    \includegraphics    [width=.49\linewidth]{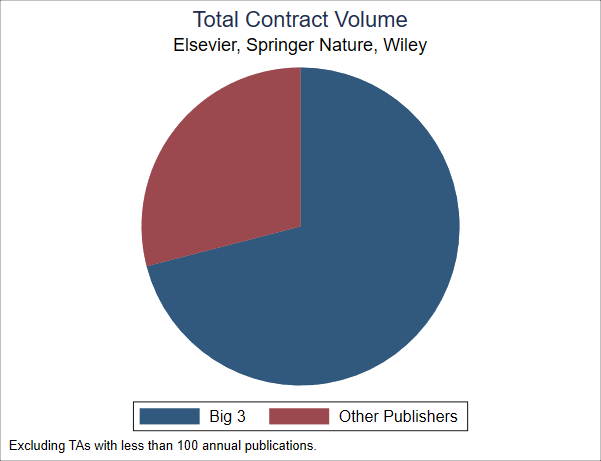}
    \caption{Aggregate Number and Volume of TAs of the `Big 3' publishers. Excluding TAs with $<10$ (upper panels)/$<100$ annual publications (lower panels).}
    \label{fig:big3_vol_small}
\end{figure}

\begin{figure}[htbp]
    \centering
    \includegraphics[width=.49\linewidth]{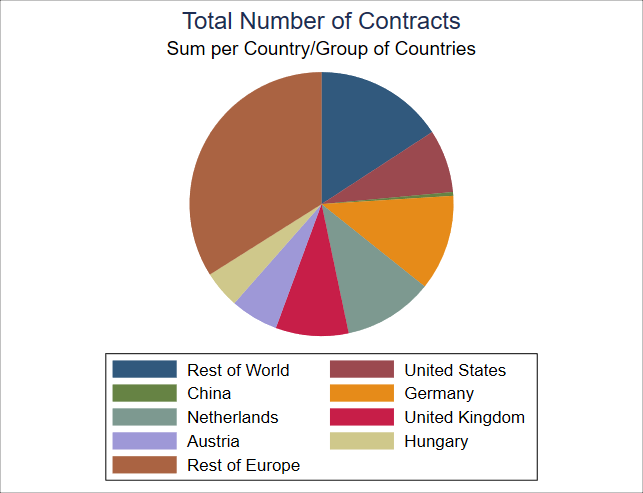}
    \includegraphics[width=.49\linewidth]{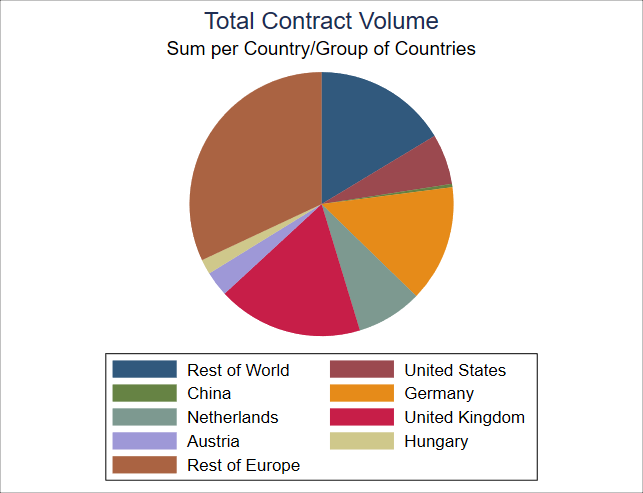}\\
    \includegraphics[width=.49\linewidth]{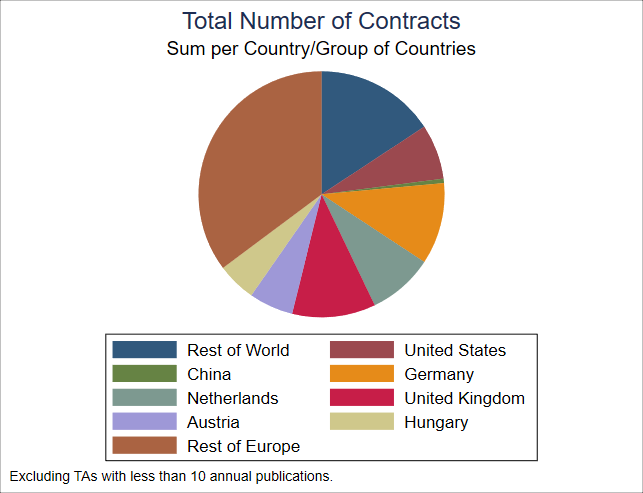}
    \includegraphics[width=.49\linewidth]{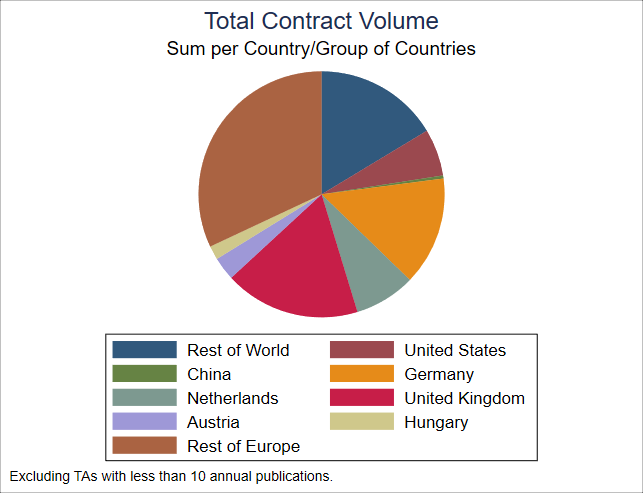}\\
    \includegraphics[width=.49\linewidth]{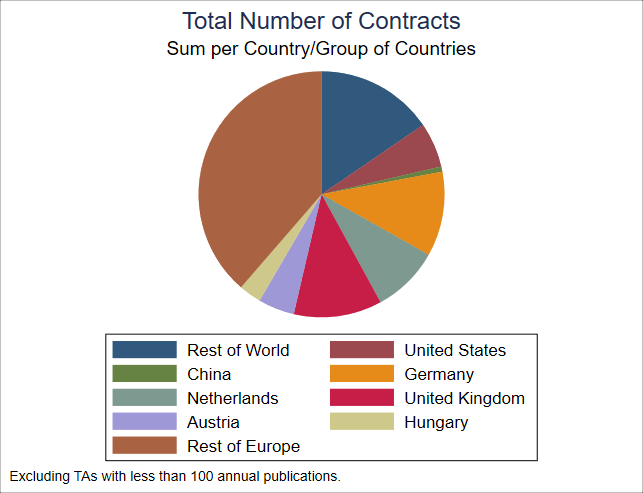}
    \includegraphics[width=.49\linewidth]{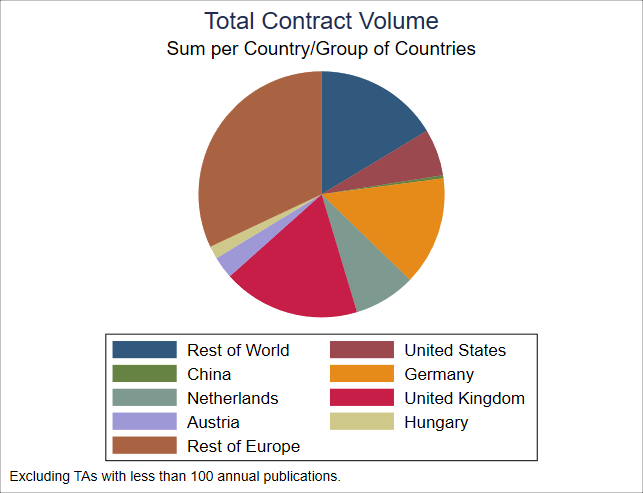}
    \caption{Aggregate Number and Volume of TAs per Country / Group of Countries. pper two panels: All TAs. Middle panels: Excluding TAs with $<10$. Bottom panels: Excluding TAs with $<100$ annual publications.}
    \label{fig:country_vol_small}
\end{figure}

\begin{table}[htbp]
\footnotesize
\begin{tabular}{lrrrr}
\toprule
Country & $\mathbbm{1}_{big\:3}=0$ & Share & $\mathbbm{1}_{big\:3}= 1$ & Share \\
\midrule
Germany       & 118 & 93,65\%  & 8   & 6,35\%  \\
Netherlands   & 109 & 92,37\%  & 9   & 7,63\%  \\
UK            & 91  & 94,79\%  & 5   & 5,21\%  \\
Austria       & 54  & 85,71\%  & 9   & 14,29\% \\
Hungary       & 40  & 81,63\%  & 9   & 18,37\% \\
Europe: Other & 336 & 82,76\%  & 70  & 17,24\% \\
China         & 5   & 100\% & 0   & 0,00\%  \\
United States & 72  & 86,75\%  & 11  & 13,25\% \\
Other         & 105 & 81,40\%  & 24  & 18,60\% \\
\midrule
Total         & 930 &          & 145 &         \\
\bottomrule
\end{tabular}
\caption{Share of TAs concluded with `Big 3' and other publishers by selected countries}
\label{tab:share_Big3_Country3}
\end{table}

\begin{table}[htbp]
\scriptsize
    \begin{center}
    \begin{tabular}{lrrr}
    \toprule
    \multicolumn{4}{c}{\textbf{\emph{Dependent variable: TA Duration}}(in logs)}\\
       Coefficient             &       \multicolumn{1}{c}{(1)}   &       \multicolumn{1}{c}{(2)}   &       \multicolumn{1}{c}{(3)}  \\
          \midrule

$\mathbbm{1}_{big\:3}$ &       0.269***&       0.272***&       0.243***\\
                    &      (0.05)   &      (0.05)   &      (0.05)   \\
Constant            &       6.657***&       6.999***&       7.256***\\
                    &      (0.04)   &      (0.00)   &      (0.12)   \\
                    \midrule
                    \multicolumn{4}{c}{\emph{Fixed Effects}} \\
                    \midrule
                    Publisher & - & - & -  \\
                    Start year & - & \checkmark & \checkmark  \\
                    Country & - & - & \checkmark \\
                    \midrule
R$^2$               &     0.034     &       0.064   &       0.284   \\
BIC                 &    1540.875   &    1570.003   &    1512.220   \\
$N$                 &        1075   &        1075   &        1075   \\
\bottomrule
    \end{tabular}\\
    {\footnotesize $* p<0.10, ^{**} p<0.05, ^{***} p<0.01$}
        \end{center}
    \caption{Relationship between contract duration and a `big 3' publisher TA. Estimation method: OLS following eq.~(\ref{eq2}). Dependent variable: log(TA-Duration). 95\% standard errors in brackets below, heteroskedasticity robust and clustered on the publisher level.}
    \label{tab:big3_length}
\end{table}

\begin{table}[htbp]
\scriptsize
    \begin{center}
    \begin{tabular}{lrrrr}
    \toprule
    \multicolumn{5}{c}{\textbf{\emph{Dependent variable: TA Volume}}(in logs)} \\
                    &       \multicolumn{1}{c}{(1)}   &       \multicolumn{1}{c}{(2)}   &       \multicolumn{1}{c}{(3)}  &       \multicolumn{1}{c}{(4)}   \\
                    \midrule
                    \midrule
                    \multicolumn{5}{c}{\emph{Excluding TAs with $<10$ annual publications}} \\
                    \midrule
log(TA-Duration)    &       1.080***&       0.687***&       0.667***&       0.439***\\
                    &      (0.21)   &      (0.13)   &      (0.13)   &      (0.15)   \\
Constant            &      -2.478*  &      -0.457   &      -0.816   &       2.243   \\
                    &      (1.24)   &      (0.90)   &      (1.19)   &      (1.43)   \\
                                        \midrule
                    \multicolumn{5}{c}{\emph{Fixed Effects}} \\
                    \midrule
                    Publisher & - & \checkmark & \checkmark & \checkmark \\
                    Start year & - & - & \checkmark & \checkmark \\
                    Country & - & - & - & \checkmark\\
                    \midrule
R$^2$               &       0.100   &       0.514   &       0.518   &       0.661   \\
BIC                 &    3253.121   &    2717.951   &    2772.277   &    2664.755   \\
N                   &         857   &         857   &         857   &         857   \\
\midrule
\midrule
\multicolumn{5}{c}{\emph{Excluding TAs with $<100$ annual publications}} \\
\midrule
log(TA-Duration)    &       0.578***&       0.434** &       0.402** &       0.221   \\
                    &      (0.19)   &      (0.17)   &      (0.17)   &      (0.16)   \\
Constant            &       2.158*  &       2.503** &       2.491** &       2.924***\\
                    &      (1.21)   &      (1.22)   &      (1.20)   &      (1.07)   \\
                                      \midrule
                    \multicolumn{5}{c}{\emph{Fixed Effects}} \\
                    \midrule
                    Publisher & - & \checkmark & \checkmark & \checkmark \\
                    Start year & - & - & \checkmark & \checkmark \\
                    Country & - & - & - & \checkmark\\
                    \midrule  
R$^2$               &       0.045   &       0.465   &       0.476   &       0.744   \\
BIC                 &    1409.260   &    1149.057   &    1188.581   &     989.286   \\
N                   &         439   &         439   &         439   &         439   \\
\bottomrule
    \end{tabular}\\
    {\footnotesize $* p<0.10, ^{**} p<0.05, ^{***} p<0.01$}
        \end{center}
    {\footnotesize Estimation method: OLS following eq.~(\ref{eq2}). Dependent variable: log(TA Volume). 95\% standard errors in brackets below, clustered on the publisher level. Regression results for the relationship between contract volume and length, excluding contracts with less than 10 (upper panel) or else less than 100 annual publications (lower panel). Excluding even larger contracts causes a lack of statistical power due to a small sample size. }
    \caption{Regression results for the relationship between TA volume and length: Excluding small contracts}
    \label{tab:big3_length2}
\end{table}


\begin{table}[ht!]
\scriptsize
    \begin{center}
    \begin{tabular}{lrrrr|r}
    \toprule
    \multicolumn{6}{c}{\textbf{\emph{Dependent variable: New TAs getting a renewal}}} \\
    \midrule
&       \multicolumn{1}{c}{(1)}   &    \multicolumn{1}{c}{(2)}   &       \multicolumn{1}{c}{(3)}  &       \multicolumn{1}{c}{(4)} & \multicolumn{1}{c}{(5)} \\
   Coefficient            & \multicolumn{1}{c}{Probit} & \multicolumn{1}{c}{Probit} & \multicolumn{1}{c}{Probit} & \multicolumn{1}{c}{Probit} & \multicolumn{1}{c}{OLS} \\
\midrule
 \multicolumn{6}{c}{\emph{Excluding TAs with $<10$ annual publications}} \\
\midrule
log(TA-Volume)      &       0.038   &       0.099*  &       0.089   &       0.175*  &       0.027** \\
                    &      (0.04)   &      (0.05)   &      (0.06)   &      (0.10)   &      (0.01)   \\
log(TA-Duration)    &      -0.793***&      -0.643***&      -0.953***&      -1.812***&      -0.287***\\
                    &      (0.12)   &      (0.15)   &      (0.23)   &      (0.33)   &      (0.06)   \\
Constant            &       4.780***&       2.805***&       3.728***&       7.104***&       3.295***\\
                    &      (0.70)   &      (0.95)   &      (1.40)   &      (1.90)   &      (0.46)   \\
\midrule
 \multicolumn{6}{c}{\textbf{\emph{Average Marginal Effect of Size}}} \\
 $\frac{\partial\:\mathbbm{1}_{Follow\:up}}{\partial\:log(Size)}$ & 0.013 & 0.032*  & 0.022 & 0.033** & 0.027** \\
 & (0.01) & (0.02) & (0.01) & (0.02) & (0.01) \\
 \midrule
 \multicolumn{6}{c}{\emph{Fixed Effects}} \\
 \midrule
 Publisher &-& \checkmark & \checkmark & \checkmark & \checkmark \\
 Start year & - & - & \checkmark & \checkmark & \checkmark\\
 Country & - & - & - & \checkmark & \checkmark\\
\midrule
Pseudo R$^2$         &       0.054   &       0.086   &       0.325   &       0.508   &  -  \\
R$^2$               & - & - &  -  &  -    &       0.577   \\
BIC                 &     731.950   &     650.513   &     484.723   &     460.349   &     512.104   \\
$N$                   &         589   &         555   &         503   &         422   &         589   \\
\midrule
\midrule
 \multicolumn{6}{c}{\emph{Excluding TAs with $<100$ annual publications}} \\
 \midrule
log(TA-Volume)      &       0.114   &       0.173*  &       0.254***&       0.189** &       0.059** \\
                    &      (0.07)   &      (0.10)   &      (0.10)   &      (0.09)   &      (0.02)   \\
log(TA-Duration)    &      -0.860***&      -0.622***&      -1.195***&      -2.388***&      -0.370***\\
                    &      (0.17)   &      (0.23)   &      (0.35)   &      (0.52)   &      (0.08)   \\
Constant            &       4.723***&       1.872   &       4.256** &      11.716***&       1.866***\\
                    &      (0.99)   &      (1.38)   &      (1.82)   &      (2.67)   &      (0.53)   \\
\midrule
\multicolumn{6}{c}{\textbf{\emph{Average Marginal Effect of Size}}} \\
 $\frac{\partial\:\mathbbm{1}_{Follow\:up}}{\partial\:log(Size)}$ & 0.038 & 0.054* &  0.062*** & 0.035** & 0.059**\\
 & (0.02) & (0.03) & (0.02) &  (0.01) & (0.02)\\
 \midrule
 \multicolumn{6}{c}{\emph{Fixed Effects}} \\
 \midrule
 Publisher & - & \checkmark & \checkmark & \checkmark & \checkmark \\
 Start year & - & - & \checkmark & \checkmark & \checkmark\\
Country & - & - & - & \checkmark & \checkmark\\
\midrule
Pseudo R$^2$         &       0.055   &       0.092   &       0.305   &       0.515   &    -           \\
R$^2$     & -  &     -   &  -   &  -  &       0.586   \\
BIC                 &     384.229   &     332.327   &     266.361   &     222.592   &     255.798   \\
$N  $                 &         314   &         289   &         258   &         204   &         314   \\
\bottomrule
    \end{tabular}\\
    {\footnotesize $* p<0.10, ^{**} p<0.05, ^{***} p<0.01$}
        \end{center}
    \caption{Regression results for the relationship between a TA getting a renewal and its size and length. Estimation method: Probit/OLS following eq.~(\ref{eq3}). Dependent variable: $\mathbbm{1}_{Renew}$. AME computed following eq.~(\ref{eq4}). 95\% standard errors in brackets below, heteroskedasticity-robust and clustered on the publisher level. Regression results for the relationship between an inital TA getting a renewal and its initial contract size, excluding contracts with less than 10 (upper panel) or else less than 100 annual publications (lower panel). Excluding even larger contracts causes a lack of statistical power due to a small sample size.}
    \label{tab:renewal2}
\end{table}

\end{doublespace}
\end{document}